\documentclass{aa}
\usepackage{graphicx}
\newcommand{\nablaa}{\nabla_{\alpha}}
\newcommand{\nablab}{\nabla_{\beta}}
\newcommand{\pad}{\partial}
\newcommand{\md}{\mbox{d}}
\newcommand{\beq}{\begin{equation}}
\newcommand{\eeq}{\end{equation}}
\newcommand{\beqn}{\begin{eqnarray}}
\newcommand{\eeqn}{\end{eqnarray}}

\begin{document}
   \title{Particle acceleration in rotating and shearing jets from AGN}

   \author{F.~M. Rieger
          \inst{1,2}
          \and
          K. Mannheim\inst{1}}

   \offprints{F.~Rieger (frieger@uni-sw.gwdg.de)}

   \institute{Institut f\"ur Theoretische Physik und Astrophysik der
              Universit\"at W\"urzburg, Am Hubland, 97074 W\"urzburg\\
              \email{mannheim@astro.uni-wuerzburg.de}
              \and
              Universit\"ats-Sternwarte G\"ottingen, Geismarlandstr. 11,
              D-37083 G\"ottingen\\
              \email{frieger@uni-sw.gwdg.de}}

   \date{Received 18 March 2002; accepted 26 September 2002}

   \abstract{
   We model the acceleration of energetic particles due to shear and
   centrifugal effects in rotating astrophysical jets. 
   The appropriate equation describing the diffusive transport of energetic 
   particles in a collisionless, rotating background flow is derived and 
   analytical steady state solutions are discussed. In particular, by 
   considering velocity profiles from rigid, over flat to Keplerian rotation, 
   the effects of centrifugal and shear acceleration of particles scattered 
   by magnetic inhomogeneities are distinguished. In the case where shear 
   acceleration dominates, it is confirmed that power law particle momentum 
   solutions $f(p) \propto p^{-(3+\alpha)}$ exist, if the mean scattering time 
   $\tau_c \propto p^{\alpha}$ is an increasing function of momentum. 
   We show that for a more complex interplay between shear and centrifugal 
   acceleration, the recovered power law momentum spectra might be 
   significantly steeper but flatten with increasing azimuthal velocity 
   due to the increasing centrifugal effects.
   The possible relevance of shear and centrifugal acceleration for the 
   observed extended emission in AGN is demonstrated for the case of the 
   jet in the quasar 3C273.
    }

   \maketitle
%
%________________________________________________________________

\section{Introduction}
    Astrophysical jets emerge from a variety of astrophysical sources,
    ranging from protostellar objects, to micro-quasars, active galactic
    nuclei (AGN) and probably Gamma Ray Bursts (GRBs). 
    Currently, the observations of jets from radio-loud AGN may be counted
    among the most interesting ones as they constitute test laboratories for 
    studying the spatial structure in relativistic jets.    
    Today there is convincing evidence that the central engine in these 
    AGN is a rotating supermassive black hole surrounded by a geometrically
    thin accretion disk, which gives rise to the formation of a pair 
    of relativistic jets.  
    The observation of superluminal motions and theoretical opacity arguments 
    indicate that the bulk plasma in these jets moves at relativistic speeds 
    along the jet axis. Models based on the assumption of a one-dimensional
    velocity structure have thus allowed useful insights into the emission 
    properties of AGN jets. However, as real jets are generally expected to 
    show a significant velocity shear perpendicular to their axes, such
    models may be adequate only for a first approximation. In particular,
    several independent arguments suggest that AGN jet flows might be
    characterized by an additional rotational velocity component:\\
    (1) The analysis of the jet energetics in extragalactic radio sources
    have revealed a remarkably universal correlation between a disk luminosity
    indicator and the bulk kinetic power in the jet (Rawlings \& 
    Saunders~\cite{rawlings91}; Celotti \& Fabian~\cite{celotti93}) and
    supported a close link between jet and disk. The successful application
    of models within the framework of a jet-disk symbiosis (Falcke \& 
    Biermann~\cite{falcke95a}; Falcke et al.~\cite{falcke95b}) indicates
    that for radio-loud objects the total jet power may approach 1/3 of the 
    disk luminosity so that a considerable amount of accretion energy, and
    hence rotational energy of the disk (cf. virial theorem), is channelled 
    into the jet leading to an efficient removal of angular momentum 
    from the disk.\\  
    (2) Observational findings, including helical motion of knots or
    periodic variabilities, seem to provide additional evidence for 
    intrinsic rotation in AGN jets (cf. Biretta~\cite{biretta} for M87;
    Camenzind \& Krockenberger~\cite{ckrocke92} for 3C273; Schramm et 
    al.~\cite{schrammetal} for 3C345).\\
    (3) From a more theoretical point of view, intrinsic jet rotation is 
    generally expected in magnetohydrodynamical (MHD) models for the formation
    and collimation of astrophysical jets (e.g. Begelman~\cite{begelman}; 
    Sauty et al.~\cite{sauty02}). In such models, intrinsic rotation with 
    speeds up to a considerable fraction of the velocity of light is a natural
    consequence of the assumption that the flow is centrifugally accelerated 
    from the accretion disk. It should be noted however, that the rotation 
    profile in the jet does not necessarily have to be disk-like, i.e. the set
    of available jet rotation profiles could be much wider and might include, 
    for example, rigid, flat and Keplerian profiles (e.g. Hanasz et 
    al.~\cite{hanasz}). 
    In particular, rigid rotation inside a well-defined light cylinder might 
    be related to foot points of the magnetic field lines concentrated near 
    the innermost stable orbit (e.g. Camenzind~\cite{camenzind}; 
    Fendt~\cite{fendt97a}), while more generally differential rotation 
    would be intuitively expected if there is an intrinsic connection between 
    jet motion and the rotating disk (cf. also Fendt~\cite{fendt97b}; 
    Lery \& Frank~\cite{leryfrank}).

    In this paper, we are interested in the influence of such rotation and
    shear profiles on the acceleration of particles in AGN jets. So far, 
    several authors have contributed to our understanding of particle 
    acceleration by shear.
    A kinetic analysis was used in the pioneering approach of
    Berezhko~(\cite{berez81}; \cite{berez82}) and Berezhko \& Krymskii
    (\cite{berezkrymskii81}). Their results showed that the particle 
    distribution function in the steady state might follow a power law 
    if the mean interval between two scattering events increases with 
    momentum according to a power law.
    Later on, particle acceleration in the diffusion approximation at a 
    gradual shear transition in the case of non-relativistic flow velocities 
    was studied independently by Earl, Jokipii \& Morfill (\cite{earletal}).
    They re-derived Parker's equation (i.e. the transport equation including 
    the well-known effects of convection, diffusion and adiabatic energy 
    changes), but also augmented it with new terms describing the viscous 
    momentum transfer and the effects of inertial drifts. 
    Jokipii \& Morfill (\cite{jokipiimorfill}) used a microscopic treatment 
    to analyse the non-relativistic particle transport in a moving, 
    scattering fluid which undergoes a step-function velocity change in the 
    direction normal to the flow. They showed that particles may gain energy
    at a rate proportional to the square of the velocity change. 
    Matching conditions in conjunction with Monte Carlo simulations for 
    shear discontinuities were derived by Jokipii, Kota \& 
    Morfill~(\cite{jokipiietal}).
    The Monte Carlo analysis was extended by Ostrowski~(1990; 1998), who also
    studied the acceleration at a sharp tangential velocity discontinuity, 
    including relativistic flow speeds. He found that only relativistic flows
    can provide conditions for efficient acceleration, resulting in a very 
    flat particle energy spectra which depends only weakly on the scattering 
    conditions.
    The relevance of such a scenario for the acceleration of particles at the 
    transition layer between AGN jets and their ambient medium was 
    stressed in recent contributions by Ostrowski (\cite{ostrowski00})
    and Stawarz \& Ostrowski~(\cite{stawarz-ostrowski}).

    The work on gradual shear acceleration by Earl, Jokipii \& Morfill 
    (\cite{earletal}) was successfully extended to the relativistic regime 
    by Webb (\cite{webb89}; \cite{webb92}). Assuming the scattering to be 
    strong enough to keep the distribution function almost isotropic in the 
    comoving frame (so that the diffusion approximation applies), he derived 
    the general relativistic diffusive particle transport equation for both 
    rotating and shearing flows.
    Subsequently, Green's formula for the relativistic diffusive particle 
    transport equation was developed by Webb, Jokipii \& Morfill 
    (\cite{webbetal94}). Applying their results to the cosmic ray transport 
    in the galaxy, they found that the acceleration of cosmic rays beyond 
    the knee by means of galactic rotation might be possible, but not to
    a sufficient extent. In its form presented however, it rather remains 
    an essentially theoretical approach which could not be easily related 
    to observations.

    In a previous contribution (Rieger \& Mannheim~\cite{riegermannheim01b})
    we developed and applied a new model, that utilizes the relativistic 
    transport theory advanced by Webb and that permits the analysis of the 
    acceleration of energetic particles in rotating and shearing AGN jets. 
    By studying velocity profiles known to be typical for such jets, 
    we obtained results indicating that the resultant particle energization 
    is in general a consequence of both centrifugal and shear effects. 
    The formation of power law particle spectra under a wide range of 
    conditions reveals the significant potential of shear and centrifugal 
    acceleration as a natural explanation for the origin of the extended, 
    continuous emission recently observed from AGN (e.g. Meisenheimer et 
    al.~\cite{meisenheimeretal97}; Jester et al. \cite{jesteretal01}). 
    In the present paper we aim to provide a more detailed investigation 
    of this model, including the derivation of the relevant particle 
    transport equation (cf. appendix~B).

    The paper is organised as follows: After a short review of the underlying
    theoretical background in part~2 (see also appendix~A), we present several 
    applications to relativistic jet flows with rigid, Keplerian and flat 
    intrinsic rotation profiles in part~3, leaving the detailed derivation of 
    the particle distribution function to appendix~C.
    Part~4 provides a discussion of our results. The observational relevance 
    of the model presented is pointed out in part~5 with reference to recent 
    observations. The paper finally closes with a short conclusion.

\section{Model background}
    By starting from the relativistic Boltzmann equation and by using both
    the differential moment equations and a simple BKG (Bhatnagar, Gross \&
    Krook) time relaxation for the scattering term, the relativistic particle 
    transport equation in the diffusion approximation was derived by 
    Webb~(\cite{webb89};~\cite{webb92}; cf. also appendix~\ref{difftransport}).
    In the underlying physical picture, it is assumed that scattering of high 
    energy particles occurs by small-scale magnetic field irregularities
    carried in a collisionless, systematically moving background flow. In each 
    scattering event the particle momentum is randomized in direction, 
    but its magnitude $p'$ is assumed to be conserved in the (local) comoving 
    flow frame, where the electric field vanishes. 
    Since the rest frame of the scattering centres is regarded to be 
    essentially that of the background flow, particles neither gain energy
    nor momentum merely by virtue of the scattering if there is no shear or 
    rotation present and the flow is not diverging.
    However, in the case of a shear in the background flow, the 
    particle momentum relative to the flow changes for a particle 
    travelling across the shear. 
    As the particle momentum in the local flow frame is preserved in the 
    subsequent scattering event, a net increase in particle momentum may occur 
    (cf. Jokipii \& Morfill~\cite{jokipiimorfill}).
    Thus, if rotation and shear is present, high energy particles, which do 
    not corotate with the flow, will sample the shear flow and may be 
    accelerated by the centrifugal and shear effects (cf. Webb et 
    al.~\cite{webbetal94}).

    For the present application, we consider a rather idealized (hollow) 
    cylindrical jet model where the plasma moves along the $z$-axis at 
    constant (relativistic) $v_z$, while its velocity component in the plane 
    perpendicular to the jet axis is purely azimuthal and characterized by the
    angular frequency $\Omega$. Using cylindrical coordinates for the position
    four vector, i.e. $x^{\alpha}=(ct,r,\phi,z)$, the metric tensor becomes 
    coordinate-dependent (i.e. $(g_{\alpha \beta})=diag\{-1,1,r^2,1\}$) and 
    for the chosen holonomic basis the considered flow four velocity may be 
    written in shortened notation as 
    \beqn\label{velocity} 
       u^{\alpha}&=& \gamma_f\,(1,0,\Omega/c,v_z/c)\,,\\
       u_{\alpha}&=& \gamma_f\,(-1,0,\Omega\,r^2/c,v_z/c)\,,
    \eeqn where the normalization 
    \beq
        \gamma_f=1/\sqrt{1-\Omega^2 r^2/c^2-v_z^2/c^2}
    \eeq
    denotes the Lorentz factor of the flow and where the angular frequency 
    may be a function of the radial coordinate, i.e. $\Omega=\Omega(r)$.\\ 
    As suggested by Webb  et al.~(\cite{webbetal94}), the resulting transport 
    equation may be cast in a more suitable form if one replaces the comoving 
    variable $p'$ by the variable $\Phi=\ln(H)$, where $H$ is given by
    \beq\label{ham}
       H=p'^0\,c\,\exp\left(-\int^r \md r'\frac{\gamma_f^2\,\Omega^2\,r'}{c^2}
       \right)\,.
    \eeq
    In the case of highly relativistic particles (with $p'^0 \simeq p'$)  
    and using an (isotropic) diffusion coefficient of the form 
    \beq
        \kappa=\kappa_0\; p'^{\alpha}\,r^{\beta}\,,
    \eeq
    we finally arrive at the steady state transport equation 
    (cf. appendix~\ref{cylinderform}) for the (isotropic) phase space 
    distribution function $f(r,z,p')$
    \beqn\label{transport-simple}
      \frac{\pad^2 f}{\pad r^2}&+&\left(\frac{1+\beta}{r}+[3+\alpha]\,
            \frac{\gamma_f^2\,\Omega^2\,r}{c^2}\right)\,\frac{\pad f}{\pad r}
      \nonumber \\
        &+&\frac{\gamma_f^4\,r^2}{5\,c^2}\,(1-v_z^2/c^2)\,
           \left(\frac{\md \Omega}{\md r}\right)^2
           \left([3+\alpha]\frac{\pad f}{\pad \Phi}+
           \frac{\pad^2 f}{\pad \Phi^2}\right)
      \nonumber\\
        &-&\frac{\gamma_f\,v_z}{\kappa}\,\frac{\pad f}{\pad z}
         + (1+\gamma_f^2\,v_z^2/c^2)\,\frac{\pad^2 f}{\pad z^2}
         =-\frac{Q}{\kappa}\,.
    \eeqn
    In general, the solution of Eq.~(\ref{transport-simple}) can be quite
    complicated. However, as we are especially interested in the azimuthal 
    effects of particle acceleration in a rotating flow further along the
    jet axis, it seems sufficient for a first approach to search for a 
    $z$-independent solution of the transport equation, i.e. we may be 
    content with an investigation of the one-dimensional Green's function, 
    which preserves much of the physics involved and which corresponds to
    the assumption of a continuous injection along the jet (cf. appendix~C).
    In this case the corresponding source term becomes
    \beq
       Q=\frac{q_0}{p_s'}\; \delta(r-r_s)\,\delta(\Phi-\Phi_s)\,, 
    \eeq describing mono-energetic injection of particles with momentum 
    $p'=p_s'$ from a cylindrical surface at $r=r_s$. Here, $\delta$ denotes 
    the Dirac delta distribution and the constant $q_0$ is defined as 
    $q_0=N_s/(8\,\pi^2\,p_s'^2\,r_s)$ with the total number $N_s$
    of injected particles. In order to solve the $z$-independent transport 
    equation, we then apply Fourier techniques and consider the Green's 
    solutions satisfying homogeneous, i.e. zero Dirichlet boundary conditions.

\section{Applications}
\subsection{Rigid rotation - no shear}
\begin{figure}[htb]
      \centering
%      \vspace*{2.0mm} 
      \includegraphics[width=7.5cm]{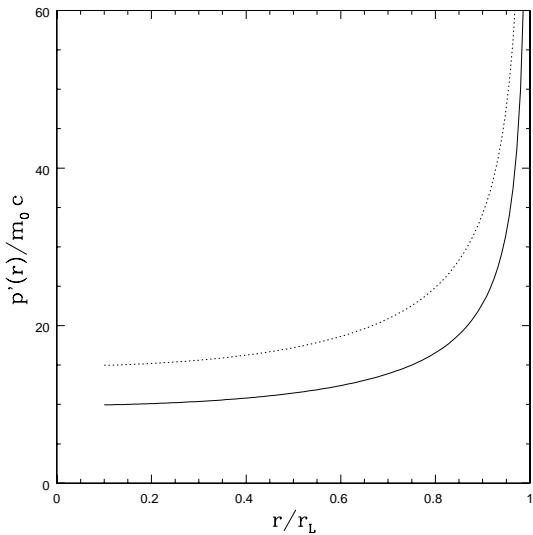}
      \caption{Particle momentum $p'$ as a function of the radial coordinate 
      in the case of rigid rotation, for particles injected at $r_s=0.1\,
      r_{\rm L}$ with initial Lorentz factor $\gamma_s=10$ (solid) and $15$ 
      (dashed). Here, $r_{\rm L}$ is defined by $r_{\rm L}=c\,
      (1-v_z^2/c^2)^{0.5}/\Omega_0$}\label{fig0}
    \end{figure}

\begin{figure}[htb]
      \centering
%      \vspace*{2.0mm} 
      \includegraphics[width=8.0cm]{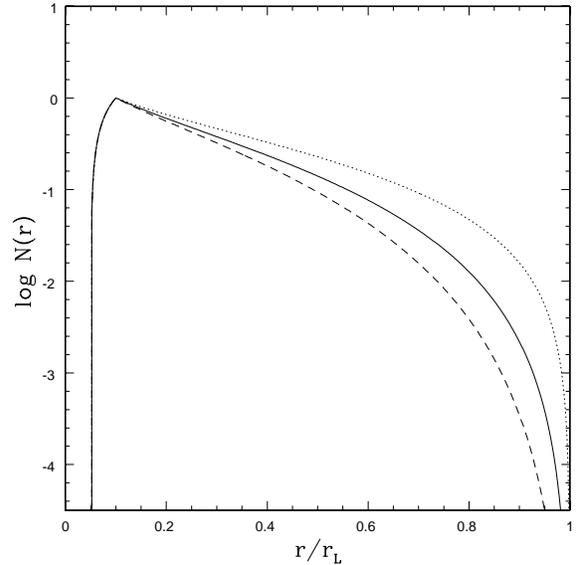}
      \caption{Spatial distribution $N(r)/N(r_{\rm s})$ for rigid rotation
      using a different energy dependence of the diffusion coefficients, i.e. 
      $\alpha=\beta=0$ (solid line), $\alpha=-2$, $\beta=-0.01$ (dotted line) 
      and $\alpha=2$, $\beta=-0.01$ (dashed line). Boundary conditions 
      $r_{\rm in}=0.05\,r_{\rm L}$, $r_{\rm s}=0.1\,r_{\rm L}$ and 
      $r_{\rm out}=0.999\,r_{\rm L}$ has been used for the 
      calculations.}\label{fig1}
    \end{figure}

   In order to gain insight into the particle transport and acceleration 
   process, let us first consider the influence of rigid rotation profiles. 
   One may associate such rotation profiles with dynamo action in the inner 
   accretion disk around a spinning black hole, which creates a jet 
   magnetosphere filled with disk plasma and rotating with the angular 
   frequency of its foot points concentrated near the innermost stable 
   orbit (e.g. Camenzind~\cite{camenzind}). 
   The acceleration of particles due to rigid rotation could represent an 
   efficient process for the production of high energy particles. 
   This was demonstrated by Rieger \& Mannheim~(\cite{riegermannheim00})
   for the centrifugal acceleration of charged test particles at the 
   base of a rigidly rotating jet magnetosphere for conditions assumed to
   prevail in AGN-type objects.  
   In such cases an upper limit for the maximal attainable Lorentz factor can
   be established, which is given by the breakdown of the bead-on-the-wire 
   approximation in the vicinity of the light cylinder. 
   There are several similarities to the application presented here:
   If we consider the transport of relativistic particles in a rigidly 
   rotating background flow (i.e. $\Omega = \Omega_0= const.$) by utilising
   the above model, shearing is absent (i.e. $\md \Omega/\md r =0$) and
   thus particle energization only occurs as a consequence of the 
   interaction with the accelerating background flow, i.e. due to the presence
   of centrifugal effects. 
   It can then be readily shown that for the case considered here, 
   Eq.~(\ref{ham}) is analogous to the Hamiltonian for a bead on a rigidly 
   rotating wire (cf. Webb et al.~\cite{webbetal94}; also Rieger \& Mannheim 
   \cite{riegermannheim01a}), i.e. 
    \beq\label{hami-rigid}
       H=\frac{p'^0\,c}{\gamma_f}\,,
   \eeq while the transport equation becomes purely spatial.
   The variable $H$ introduced above, could thus be regarded as describing the 
   balance between the centrifugal force and the inertia of the particle in 
   the comoving frame.
   By means of Noether's theorem, $H$ could be shown to be a constant of 
   motion, and hence Eq.~(\ref{hami-rigid}) indicates that the ratio of 
   particle to flow Lorentz factor is fixed. The comoving particle momentum 
   $p'$ can then be simply expressed as a function of the 
   radial coordinate $r$, i.e. $p' = p'(r)$ (cf. Fig.~\ref{fig0}).\\
   We may easily solve the resultant (spatial) transport equation analytically,
   assuming homogeneous boundary conditions at the jet inner radius 
   $r_{\rm in}$ and its relevant outer radius $r_{\rm out}$. In the present 
   case, one may choose the size of the inner radius $r_{\rm in}$ to be of the
   order of the radius of the innermost stable orbit around a rotating black 
   hole, while the outer radius $r_{\rm out}$ is at any rate physically 
   constrained to be smaller than the light cylinder radius (note that for 
   relativistic MHD winds, the Alfv\'{e}n point may be close to the
   light cylinder).
   Results concerning the particle transport are presented in Figs.~\ref{fig0}
   and \ref{fig1}. 
   They indicate that the energetic particles injected at $r_s$ will gain 
   energy by being transported in the rigidly rotating background flow, 
   i.e. due to the impact of centrifugal acceleration the particle momentum 
   increases with position as expected. The precise evolution in the 
   immediate vicinity of the light cylinder should however be treated with 
   some caution due to the limitations imposed by the applied diffusion 
   approach.
   Fig.~\ref{fig1} reveals a decrease in efficiency if the time $\tau_c$ 
   between two collisions increases with momentum (in case of a very weak 
   spatial dependence), i.e. if scattering occurs less rapidly for the higher 
   than for the lower energy particles.

%-------------------------------------------------------------------------%
%-------------------------------------------------------------------------%

\subsection{Keplerian rotation - shear dominance}
\begin{figure}[htb]
         \centering
%         \vspace*{2.0mm}
         \includegraphics[width=8.0cm]{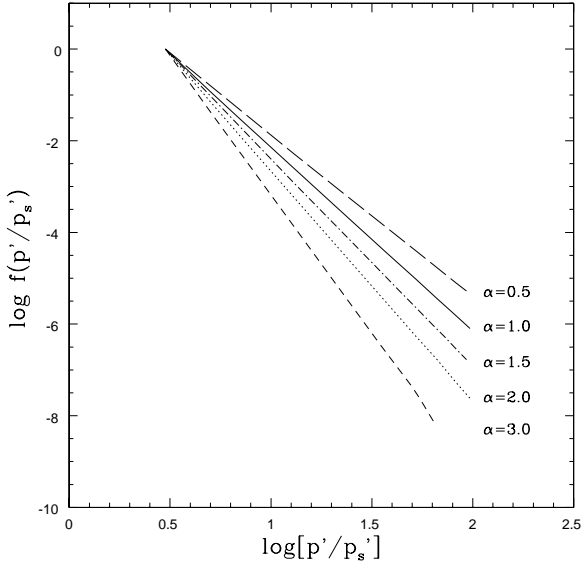}
         \caption{The momentum-dependence of the (normalized) distribution
         function $f$ for Keplerian rotation using $\beta=0$, calculated for 
         $\alpha=0.5, 1.0, 1.5, 2.0, 3.0$ at fixed $r=40\,r_{\rm ms}$.  
         Boundary and injection conditions have been specified as 
         $r_{\rm in}=10\,r_{\rm ms}$, $r_{\rm out}=1000\,r_{\rm ms}$, 
         $r_{\rm s}=20 \,r_{\rm ms}$, where $r_{\rm ms}$ is given by 
         as $r_{\rm ms}=G\,M/(c^2-v_z^2)$.}\label{keplerplot}
     \end{figure}

     Let us now consider the acceleration of particles due to a rotating 
     background flow with keplerian rotation profile $\Omega(r)=k\,r^{-3/2}$, 
     where $k=\sqrt{G\,M}$.
     Again, such flow profiles might be associated with jets or disk winds 
     originating from the accretion disk around the black hole and dragging 
     the Keplerian disk rotation with them. Hence, for reasons of 
     self-similarity, keplerian rotation may be intuitively regarded as one 
     of the most characteristical descriptions with respect to the velocity 
     profile of intrinsically rotating flows. 
     For example, Lery \& Frank~(\cite{leryfrank}) recently investigated 
     the structure and stability of astrophysical jets including Keplerian 
     rotation in the outermost part of the outflow and rigid rotation close 
     to its axis (cf. also Hanasz et al.~\cite{hanasz}). 
     They also studied the application to non-relativistic outflows from 
     young stellar objects. One may thus eventually consider a simple model 
     where particles, accelerated in a rigidly rotating flow, are subsequently
     injected into a Keplerian rotating flow.\\
     Generally, if we consider Keplerian rotation, both shear and centrifugal 
     effects are present. For non-relativistic rotation, analytical solutions 
     of the Fourier transformed transport equation are given in the 
     appendix~C.2. in terms of the confluent hypergeometrical functions. 
     Such solutions should be appropriate for the outer jet regions. As the
     relative strength of the contribution by shear to that of centrifugal 
     energization evolves with $r$, shear effects will eventually dominate 
     over centrifugal acceleration (see discussion). 
     This is particularly illustrated in Fig.~\ref{keplerplot}, where we 
     have plotted the logarithm of the (normalized) particle distribution 
     function $f$ above $p'/p_s' \geq 3$ for $\beta=0$ (i.e. allowing no 
     spatial dependence of the diffusion coefficient) and different 
     momentum dependence of $\kappa$. 
%--------------------------------------------------------------------------%
      Most interestingly, well-developed power law momentum spectra are 
      recovered, which suggest a spectral slope linearly related to
      the momentum index of the diffusion coefficient, so that we have 
      $f \propto p'^{-(3+\alpha)}$. This excellently confirms the results
      previously derived by Berezhko~\&~Krymskii~(\cite{berezkrymskii81}).
      For a collisionless plasma with a simple shear flow $U(y)\, \vec{e}_x$, 
      and by using a simple BGK term, they found the shear acceleration to 
      give rise to a power-law momentum spectrum for the steady state 
      comoving particle distribution $n(r,p') \propto p'^{\,2}\,
      f \propto p'^{-(1+\alpha)}$, if the collision time $\tau_c$ depends on 
      momentum as $\tau_c \propto p'^{\,\alpha}$ and $\alpha >0$. 
      However, if the momentum index $\alpha$ is smaller than zero, i.e. 
      $\alpha <0$, an exponential spectrum may developed.

\subsection{Flat rotation - interplay between shear and centrifugal effects}
\begin{figure}[htb]
       \centering
       \vspace*{2.0mm}
       \includegraphics[width=8.0cm]{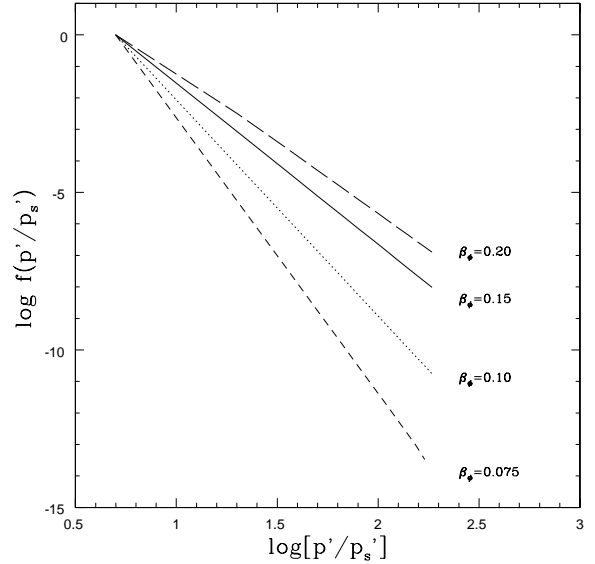}
       \caption{The momentum-dependence of the (normalized) distribution 
       function $f(r,p')$ for flat rotation, calculated for 
       $r_{\rm in}/r_{\rm out}=0.02$, $r_s/r_{\rm out}=0.04$ at position 
       $r/r_{\rm out}=0.2$. Chosen azimuthal velocities are 
       $v_{\phi f}/c=0.075$ (short-dashed), $0.10$ (dotted),$0.15$ (solid), 
       $0.20$ (long-dashed).}\label{lowfastvergleich1.eps}
      \end{figure}

      In the case of flat rotation (i.e. $\Omega=\Omega_0\,r_0/r=v_{\phi f}/r$)
      one may easily investigate a more complex interplay between shear and 
      centrifugal effects. For in this case, the relative strength of shear
      to centrifugal effects is independent of the radial coordinate and the
      general solution of the fourier-transformed equation could be cast in 
      simple analytical terms allowing basic inverse Fourier integration  
      to be done, using homogeneous boundary conditions at $r_{\rm in}$ and 
      $r_{\rm out}$. 
      For the present application, we have considered typical jet flows 
      with relativistic $v_z/c=0.95$ and different azimuthal velocities 
      $v_ {\phi f}$ (i.e. for a range of $\gamma_f \sim (3-4)$). The
      corresponding results are plotted in Fig.~\ref{lowfastvergleich1.eps} 
      using a constant diffusion coefficient (i.e. $\alpha=\beta=0$). 
      Again, the calculated distribution functions reveal a powerlaw-type 
      momentum dependence, where for low azimuthal velocities very steep 
      momentum spectra are recovered, i.e. the momentum exponents are found to
      be near by $-8.8$ (for $\beta_{\phi}=0.075$), $-6.9$ (for $\beta_{\phi}=
      0.10$), $-5.1$ (for $\beta_{\phi}=0.15$) and $-4.4$ (for $\beta_{\phi}=
      0.20$). 
      The observed flattening of the spectra with increasing azimuthal 
      velocities is indicative of the increasing impact of centrifugal 
      effects. 

\section{Discussion}
      The applications presented so far differ in their relative contributions 
      from shear and centrifugal effects. We may estimate the relative 
      strength of these two more transparently by comparing their 
      corresponding acceleration coefficients: 
      For the shear acceleration coefficient we have 
      \beq
         <\dot{p}>_s=\frac{1}{p'^2}\frac{\pad}{\pad p'}(p'^4\,\tau_c\,\Gamma)
                    =(4+\alpha)\,p'\,\frac{\lambda}{v'}\,\Gamma\, 
      \eeq where $\Gamma$ denotes the viscous energization coefficient
      given by Eq.~(\ref{energization-coefficient}) and $\tau_c \propto
      \tau_0(r)\,p'^{\alpha}$ with $\tau_c=\lambda/v'$. On the other hand,
      centrifugal acceleration may be described by the coefficient
      \beq
         <\dot{p}>_c=\frac{1}{p'^2}\frac{\pad}{\pad p'}
                     \left(\kappa\,\, (p'^0)^4\,\gamma_f^4\,
                     \frac{\Omega^4\,r^2}{c^4}\right)\,,
      \eeq which for highly relativistic particles ($p'\simeq 
      p'^0$) and $\kappa=v'^2\,\tau_c/3$ reduces to
      \beq
         <\dot{p}>_c=\frac{4+\alpha}{3}\, p'\frac{\lambda}{v'}\,
                     \gamma_f^4\,\frac{\Omega^4\,r^2}{c^4}\,.
      \eeq
      Hence, for the ratio $c_s= <\dot{p}>_c/<\dot{p}>_s$
      of centrifugal to shear acceleration we finally arrive at
      \beqn
        c_s \simeq \left\{ \begin{array}{ll}
                  2.2\, r_{ms}/r & \mbox{for keplerian rotation}\\ \\ 
                  5\, v_{\phi f}^2/(c^2-v_z^2) & \mbox{for flat rotation}
                  \end{array} \right. 
      \eeqn
      For the case of Keplerian rotation and for the parameters
      chosen above, the relative strength becomes $c_s \simeq 0.05$,
      i.e. the shear effects essentially dominate over those of centrifugal
      ones.
      On the other hand, in the case of galactic rotation we have $c_s 
      \simeq (0.3-2)$, which allows a complex interplay between centrifugal 
      and shear acceleration with increasing azimuthal velocity.

      Our approach utilises the relativistic diffusive particle 
      transport theory as advanced by Webb~(\cite{webb89}) and Webb et 
      al.~(\cite{webbetal94}) and thus assumes the diffusion approximation to 
      be valid, i.e. the deviation of the particle distribution from isotropy 
      to be small. In a strict sense, this requires the particle mean free
      path $|c\,\tau_c|$ to be much smaller than both, the typical length 
      scale for the evolution of the mean (momentum-averaged) distribution 
      function and the typical length scale of variation for the background 
      flow. Our conclusions are therefore of restricted applicability if 
      highly anisotropic distributions are expected as, for example, near 
      ultra-relativistic shocks (cf. Kirk \& Schneider~\cite{kirkschneider87}; 
      Kirk \& Webb~\cite{kirkwebb88}).

      In the application presented here, energy changes as a result of 
      radiative (e.g. synchrotron) losses or second-order Fermi effects due to
      stochastic motions of the scattering centres have not been considered. 
      One expects the inclusion of radiative losses to introduce an upper 
      bound to the possible particle energy at the point where acceleration 
      is balanced by losses, thus leading to a cut-off in the particle 
      momentum spectrum.
      On the other hand, the inclusion of second-order Fermi effects would 
      give an additional diffusion flux in momentum space. It may formally be 
      taken into account by a more careful treatment of the scattering term in
      the Boltzmann equation. For non-relativistic jet flows, a numerical 
      study of second-order Fermi acceleration has recently been given by 
      Manolakou et al.~(\cite{manolakouetal1999}). Our present omission of 
      second-order Fermi acceleration in the relativistic transport equation
      appears justifiable for the cases where the typical random velocities 
      of the scattering centres (as measured in the comoving frame, i.e. 
      relative to the flow) are smaller than a product of the radial
      particle mean free path times the rotational flow velocity gradient. 
      However, one should note, that estimating the effects of second-order 
      Fermi acceleration for the case of flat rotation, for example, clearly
      show them to be of increasing relevance for decreasing azimuthal flow 
      velocities. A more detailed analysis will thus be given in a subsequent 
      publication, while the purpose of the present model is confined to
      the analysis of steady-state solutions and the essential physical 
      features of shear and centrifugal acceleration.

      So far, shear-type acceleration processes in the context of AGN jets
      have been investigated by Subramanian et al.~(\cite{subraetal99}) and
      Ostrowski~(\cite{ostrowski98}, \cite{ostrowski00}):
      By following the road suggested by Katz~(\cite{katz91}), who considered 
      the particle acceleration in a low density corona due to flux tubes 
      anchored in a keplerian accretion disk, Subramanian et 
      al.~(\cite{subraetal99}) investigated the acceleration of protons driven
      by the shear of the underlying Keplerian accretion disk. They 
      demonstrated that the shear acceleration may transfer the energy 
      required for powering the jet and showed the shear to dominate over 
      second-order Fermi acceleration. However, their model does not deal 
      with the acceleration of particles emitting high energy radiation, but 
      is basically confined to the bulk acceleration of the jet flow up to 
      asymptotic Lorentz factors of $\sim 10$.
      On the other hand, Ostrowski~(\cite{ostrowski98},~\cite{ostrowski00}) 
      examined the acceleration of cosmic ray particles at a sharp tangential 
      flow discontinuity. He demonstrated that both, the acceleration to
      very high energies as well as the production of flat particle spectra 
      are possible. In order for this model to apply efficiently, one 
      requires several conditions to be satisfied, including the presence of 
      a relativistic velocity difference and a thin (not extended) boundary, 
      as well as a sufficient amount of turbulence on both sides of the 
      boundary. It seems however that such conditions might be realized, for 
      example, at the jet side boundary layer in powerful FR~II radio 
      sources.
 
      With respect to shear acceleration, our approach can thus be regarded
      as complementary to the one developed by Ostrowski. While our analysis 
      considers the influence of a gradual (azimuthal) shear profile in the 
      jet interior, the analysis by Ostrowski deals with a sharp tangential 
      shear profile at the jet boundary layer. Generally, one expects both 
      processes to occur and to contribute to the production of high energy 
      particles, their relative contributions being dependent on the 
      specific conditions realized in the jet interior and its boundary.

\section{Observational Relevance}
       The present results may serve as an instructive example revealing 
       the significant potential of shear and centrifugal acceleration in 
       the jets of AGN. 
       Since intrinsic jet rotation is typically expected in the AGN setting, 
       the proposed mechanism might operate in a natural manner over 
       a wide range. This suggests that particle acceleration in a rotating 
       and shearing background flow could be of particular relevance for 
       an explanation of the continuous optical emission observed from several 
       AGN jets (e.g. for 3C273; M87; PKS 0521-365; cf. 
       Meisenheimer et al.~\cite{meisenheimeretal97}, Jester et 
       al.~\cite{jesteretal01},~\cite{jesteretal02}), as
       in contrast to shock acceleration, this mechanism generally would 
       not be constrained to localized regions. Recent observations in fact 
       indicate that the radiating particles might be widely distributed 
       so that the optical emission from radio jets for example, is not 
       confined to bright individual knots (cf. Meisenheimer et 
       al.~\cite{meisenheimeretal96}, \cite{meisenheimeretal97}).
       In order to explain the apparent absence of strong radiative cooling 
       between knots, Meisenheimer, Yates \& R\"oser~(\cite{meisenheimeretal97})
       suggested the operation of an extended so-called "jet-like" acceleration
       mechanism, which was thought to be associated with velocity shear and 
       expected to contribute in addition to standard diffusive shock 
       acceleration.
       Meanwhile, the need of such an extended acceleration mechanism is 
       strengthened by recent, high-resolution HST observations of the jet 
       in the 3C273 (Jester et al.~\cite{jesteretal01},~\cite{jesteretal02}). 
       Classified as a blazar source, 3C273 is known as the brightest and 
       nearest (z=0.158) quasar, with a jet extending up to several tens of 
       kpc. The absence of a detectable counter-jet suggests bulk relativistic 
       motion even up to kpc-scales, a conclusion also supported by recent 
       models for the X-ray emission observed with Chandra (cf. Sambruna et 
       al.~\cite{sambruna01}). 
       The observations performed by Jester and collaborators have revealed 
       an extremely well-collimated jet, where the optical spectral index 
       varies only smoothly along the jet indicating only smooth variations 
       of physical conditions. Their phenomenological analysis yields 
       an optical spectral flux index of $\sim -(0.65-0.8)$ and provides
       strong evidence for an underlying, universal powerlaw electron 
       distribution, which is maintained almost entirely throughout the jet. 
       If in the context of shear and centrifugal acceleration, the particle 
       energy is locally dissipated by synchrotron radiation, the spectral 
       emissivity $j_{\rm \nu}$ for a power-law particle number density 
       distribution $n \propto p'^2\,f \propto p'^{-\delta}$ is given by 
       $j_{\rm \nu}\propto \nu^{-s}$ with $s=(\delta-1)/2$.
       Thus, for $s=(0.65-0.8)$ we require $\delta=(2.3-2.6)$, which for 
       example, may be realized for the case of a constant diffusion 
       coefficient and high flat rotation or for the case of a simple 
       shear flow with moderate momentum-dependent diffusion.

\section{Conclusion}
       Observational and theoretical arguments suggest that astrophysical
       jets should exhibit intrinsic rotation of material perpendicular to 
       the jet axis. Motivated by such arguments, we have proposed and 
       analysed a basic model for the acceleration of energetic particles 
       by centrifugal and viscous shear effects which is applicable to 
       relativistic, intrinsically rotating jet flows. 
       The results obtained indicate that particle acceleration in rotating 
       jets might represent an active mechanism for the production of the 
       synchrotron-emitting high energy particles by giving rise to 
       power law particle momentum distributions over a wide range of 
       conditions. Due to its anticipated non-localized operation, such 
       a mechanism is expected to be of particular relevance for the 
       explanation of the recently observed extended emission in the jets 
       from AGN. Moreover, as a general mechanism this process is able
       to provide high energy particles, which are required as seed
       particles for the (localised) first-order Fermi acceleration widely 
       believed to occur in relativistic jets (e.g. Drury~\cite{drury83}; 
       Kirk \& Duffy~\cite{kirkduffy99}). Particularly, in the beginning 
       the injection of seed particles could naturally occur at the base 
       of the jet due to centrifugal acceleration (cf. Rieger \& 
       Mannheim~\cite{riegermannheim00}).

\begin{acknowledgements}
      We would like to thank C.~Hettlage and D.~Rieger for a careful reading 
      of the manuscript. Useful discussions with S.~Jester and M.~Ostrowski 
      and helpful comments by an anonymous referee are gratefully 
      acknowledged. F.M.R. acknowledges support from the German
      {\emph{Deut\-sche For\-schungs\-ge\-mein\-schaft (DFG)}}, 
      project number MA 1545/2-2. 
\end{acknowledgements}

\appendix
\section{The general steady state transport in the diffusion 
         approximation}\label{difftransport}
    In the case where we are concerned with the scattering of particles in
    relativistic bulk flows, it has been typically found useful to evaluate  
    the scattering operator in the local Lorentz frame in which the fluid is 
    at rest, i.e. in the so-called comoving frame $K'$ 
    (e.g. Webb~\cite{webb85}; Riffert~\cite{riffert86}; Kirk et 
    al.~\cite{kirketal88}).\footnote{Note, that in the following, quantities 
    which are measured relative to $K'$ are labelled with a prime 
    superscript.} 
    For in this frame $K'$, a simple form of the scattering operator could be 
    applied if one assumes, as in the present approach, the rest frame of 
    the scattering centres to be essentially that of the background flow. 
    Quantities, which are operated upon the scattering operator, e.g. the 
    momentum, might then conveniently be evaluated in this comoving frame, 
    while the time and space coordinates are still measured in the laboratory 
    frame $K$ characterized by its metric tensor $g_{\alpha\,\beta}$. 
    We would like to note however, that $K'$ will in general be a 
    non-inertial coordinate system (i.e. an accelerated frame) and 
    therefore the related connection coefficients will not vanish. 
    With reference to the considered particle transport, the covariant 
    form of the Boltzmann equation is thus required, which may be achieved 
    by replacing the ordinary (partial) space-time derivatives by their 
     covariant derivatives.\\
     Now, by starting from the relativistic Boltzmann equation and
     by using a perturbation solution of the moment equations in the 
     diffusion approximation, i.e. by assuming the deviation of the 
     particle distribution from isotropy in the comoving frame to be small,
     Webb~(\cite{webb89}) has derived the general equation describing steady
     state particle transport in relativistic rotating and shearing flows. 
     Following Eq.(4.4) of Webb~(\cite{webb89}), the special relativistic 
     diffusive particle transport equation for the isotropic, mean scattering 
     frame distribution (averaged over all momentum directions) 
     $f_0(x^{\alpha}, p')=<f>$ may be written as
     \beqn\label{webb-app}
       \frac{1}{p'^2} \frac{\pad}{\pad p'}
        \left[-\frac{p'^3\,c}{3}\,f_0'\;
         \nablab\,u^{\beta} - p'\,(p'^{\,0})^2\,
         \dot{u}_{\,\alpha}\,q^{\,\alpha}
         - \Gamma\,p'^4\,\tau_c\,\frac{\pad f_0'}{\pad p'}\right] 
         \nonumber \\ 
          + \nablaa \left(c\,u^{\,\alpha}\,f_0'+ q^{\,\alpha}\right)= 0\,.
         \nonumber\\
     \eeqn with $x^{\alpha}$ the position four vector in the laboratory 
     frame K, where the background plasma is in motion with four velocity 
     $u^{\,\alpha}$, and $p'=m'\,v'$ the (magnitude of the) particle 
     momentum as measured in the local (comoving) fluid frame $K'$. Note,
     that in order to simplify matters, the subscript $0$ is omitted in 
     the following characterisations of the isotropic part of the particle
     distribution function, i.e. $f$ now denotes the isotropic part!
     The total particle energy and momentum in the frame $K'$ may be written 
     as
     \beq
      E'=p'^{\,0}\,c \quad {\rm and} \quad p'=
      \sqrt{(p'^{\,0})^2-m_0^2\,c^2}\,, 
     \eeq respectively, with $m_0$ the rest mass of the particle and 
     $c$ the speed of light.\\ 
     The terms in the first line of Eq.~(\ref{webb-app}) represent particle
     energy changes due to adiabatic expansion or compression of the flow 
     (i.e. the term proportional to the fluid four divergence 
     $\nablab\,u^{\beta}$), due to shear energization (i.e. the term 
     involving $\Gamma$) and due to the fact that $K'$ is an accelerated 
     frame (i.e. the term $\propto \dot{u}_{\,\alpha}$, cf. also
     Webb~\cite{webb85}). 
     The second line gives the effects of diffusion and convection. 
     In Eq.~(\ref{webb-app}), $\nabla_{\alpha}$ denotes the covariant 
     derivative while $q^{\,\alpha}$ denotes the heat flux. This heat flux 
     contains a diffusive particle current plus a relativistic heat  
     inertial term $\propto \dot{u}_{\beta}$ and is given by
     \beq\label{heat-app}
        q^{\,\alpha}=-\kappa^{\,\alpha\,\beta}\,
        \left(\nablab\,f_0'-\dot{u}_{\beta}
        \frac{(p'^{\,0})^2}{p'}\,\frac{\pad f}{\pad p'}\right)\,.
     \eeq 
     As shown by Webb~(\cite{webb89}), Eq.~(\ref{webb-app}) could be regarded
     as the relativistic generalization of the non-relativistic particle 
     transport equation first derived by Earl, Jokipii \& 
     Morfill~(\cite{earletal}).\\    
     The acceleration four vector $\dot{u}_{\alpha}$ of the comoving 
     (or scattering) frame in Eqs.~(\ref{webb-app}) and (\ref{heat-app}), 
     is defined by
     \beq\label{accel-app} 
       \dot{u}_{\alpha}=u^{\beta}\,\nablab\,u_{\alpha}\,.
     \eeq   
     The fluid energization coefficient $\Gamma$ in Eq.~(\ref{webb-app})
     represents energy changes due to viscosity. Since the acceleration 
     of particles draws energy from the fluid flow field, one expects on the 
     other hand the flow to be influenced by the presence of these 
     particles. As was shown, for example, by Earl, Jokipii \& 
     Morfill~(\cite{earletal}) and Katz~(\cite{katz91}), the resultant 
     dynamical effect on the flow could be modelled by means of an (induced) 
     viscosity coefficient. If one considers
     the strong scattering limit, i.e. the case where $\omega\,\tau_c \ll 1$, 
     with $\omega$ the gyrofrequency of the particle in the scattering frame
     (i.e. $\omega=q\,B'\,c/p'^0$), $\tau_c=1/\nu_c$ the mean time 
     interval between two scattering events and $\nu_c$ the collision 
     frequency, this fluid energization coefficient could be written as 
     (Webb~\cite{webb89}, 
     Eq.~34) 
     \beq\label{viscous-app}
       \Gamma =\frac{c^2}{30}\,\sigma_{\alpha\,\beta}\,
       \sigma^{\alpha\,\beta}\,,
     \eeq where $\sigma_{\,\alpha\,\beta}$ is the (covariant) fluid shear 
     tensor given by 
     \beqn\label{shear-app}
       \sigma_{\,\alpha\,\beta}&=&\nablaa u_{\beta}+\nablab u_{\alpha}
                 +\dot{u}_{\alpha} u_{\beta}+\dot{u}_{\beta} u_{\alpha} 
                 \nonumber\\ 
                 &+&\frac{2}{3}\left(g_{\,\alpha\,\beta}+u_{\alpha}\,u_{\beta}
                  \right)\,\nabla_{\delta} u^{\delta}\,,
     \eeqn with $g_{\alpha \beta}$ the (covariant) metric tensor.
     Additionally, for the strong scattering limit the spatial diffusion 
     tensor $\kappa^{\alpha \beta}$ reduces to a simple form given by
     \beq\label{diffusion-coefficient} 
     \kappa^{\alpha \beta}=\kappa\,(g^{\alpha \beta}+u^{\alpha}\,
     u^{\beta})\,, \quad {\rm with}\;\; \kappa= v'^2\,\tau_c/3\,,
     \eeq the isotropic diffusion coefficient and $v'$ the comoving particle 
     speed.\\

\section{Derivation of the steady state diffusive particle 
         transport equation in cylindrical coordinates}\label{cylinderform}
         Consider for the present purpose a cylindrical jet model where the 
         plasma moves along the $z$-axis at constant (relativistic) $v_z$ 
         while the perpendicular velocity component is purely azimuthal, i.e. 
         characterized by the angular frequency $\Omega$, in which case it 
         proves useful to apply cylindrical coordinates $x^{\alpha}=
         (ct,r,\phi,z)$. One may then choose a set of (non-normalized) 
         holonomic basis vectors $\{{\bf{e_{\alpha}}}, \alpha=0,1,2,3\}$ with
         ${\bf{e_{\alpha}}}=\pad {\bf{x}}/\pad x^{\alpha}$, which determine a 
         $1-$form basis $\{{\bf{e^{\alpha}}}\}$, known as its dual basis.
         For cylindrical coordinates $x^{\alpha}$ the metric tensor 
         $g_{\alpha \beta}$ becomes coordinate-dependent, i.e. for the 
         covariant metric tensor we have
         \beq
              (g_{\alpha \beta})=diag\{-1,1,r^2,1\}\,,
         \eeq while for the contravariant counterpart is consequently given by 
         $(g^{\alpha \beta})=diag\{-1,1,1/r^2,1\}$. In particular, all partial 
         derivatives of the metric coefficients vanish, except for the 
         $g_{22}$ coefficient, and therefore all connection coefficients or 
         Christoffel symbols of second order vanish, except for         
         \beq
          \Gamma^1_{22}=-r\,, \quad \Gamma^2_{21}=\Gamma^2_{12}=\frac{1}{r}\,.
         \eeq   
         Using the chosen holonomic basis we consider a simple plasma flow 
         where the four velocity could be written in coordinate form as 
         $u^{\alpha}{\bf{e_{\alpha}}}
         =\gamma\,{\bf{e}_0}+(\gamma\,\Omega/c)\,{\bf{e}_2}+(\gamma\,v_z/c)\, 
         {\bf{e}_3}$ and $u_{\alpha}{\bf{e}^{\alpha}}=-\gamma\,{\bf{e^0}}+
         (\gamma\,r^2\,\Omega/c)\,{\bf{e}^2}+(\gamma\,v_z/c)\,{\bf{e}^3}$, 
         respectively, or in shortened notation as 
         \beqn\label{velocity-app} 
            u^{\alpha}&=& \gamma\,(1,0,\Omega/c,v_z/c)\,,\\
            u_{\alpha}&=& \gamma\,(-1,0,\Omega\,r^2/c,v_z/c)\,,
         \eeqn where the normalization 
         \beq
            \gamma=\frac{1}{\sqrt{1-\Omega^2 r^2/c^2-v_z^2/c^2}}
         \eeq 
         denotes the Lorentz factor of the flow and where the angular 
         frequency may be selected to be a function of the radial coordinate, 
         i.e. $\Omega=\Omega(r)$.\\

         Generally, for a contravariant four vector $A^{\alpha}$ the covariant 
         derivative is given by  
         \beq
          A^{\alpha}_{\;\;||\beta}=\frac{\pad A^{\alpha}}{\pad x^{\beta}}+
                            \Gamma^{\alpha}_{\beta \mu}\,A^{\mu}\,,
         \eeq while for the covariant derivative of a covariant four vector 
         $A_{\alpha}$ one has
         \beq  
            A_{\alpha||\beta}=\frac{\pad A_{\alpha}}{\pad x^{\beta}}-
                            \Gamma^{\mu}_{\alpha \beta}\,A_{\mu}\,.
         \eeq
         Hence, for the assumed four velocity Eq.~(\ref{velocity-app}), the 
         fluid four divergence becomes zero, i.e. $\nablab\, u^{\beta} =0$, 
         while the fluid four acceleration Eq.~(\ref{accel-app}) reduces to 
         \beq\label{four-acceleration-cylindrical}
            \dot{u}_{\alpha}\,{\bf{e^{\alpha}}}=-u^2\,(\Gamma^2_{21} u_2)\,
            {\bf{e}^1}=-(\gamma^2\,\Omega^2\,r/c^2)\,{\bf{e}^1}\,.
         \eeq
         For the components of the shear tensor Eq.~(\ref{shear-app}) we may 
         then derive the following relations
         \beqn\label{shear-components1}
             \sigma_{01}&=&\sigma_{10} =-(\gamma^3\,r^2/c^2)\,\Omega\,
                                 \frac{d\Omega}{dr}\,\nonumber\\
             \sigma_{00}&=&\sigma_{11} =\sigma_{22}=\sigma_{33}=\sigma_{02}
                  =\sigma_{20}=\sigma_{23}=\sigma_{32}=0\,\nonumber\\
             \sigma_{12}&=&\sigma_{21} =(\gamma^3\,r^2/c)\,\frac{d\Omega}{dr}\,
                                 (1-v_z^2/c^2)\,\nonumber\\
             \sigma_{13}&=&\sigma_{31} =(\gamma^3\,r^2\,v_z/c^3)\,\Omega\,
                                 \frac{d\Omega}{dr}\,\,.
         \eeqn The viscous energization coefficient  
         Eq.~(\ref{viscous-app}) then becomes
         \beq\label{energization-coefficient}
          \Gamma=\frac{1}{15}\,\gamma^4\,r^2\,\left(\frac{d\Omega}{dr}\right)^2
                    (1-v_z^2/c^2)\,,
         \eeq noting that $\sigma^{01}=-\sigma_{01}$, $\sigma^{12}=
         \sigma_{12}/r^2$ and $\sigma^{13}=\sigma_{13}$.\\ 

         As the fluid four divergence vanishes, the first term in brackets 
         of Eq.~(\ref{webb-app}) becomes zero, while for the second term in 
         brackets one finds     
         \beq\label{radial-cylindrical}
             \dot{u}_{\alpha}\,q^{\alpha}= \dot{u}_1\,q^1\,,
         \eeq with the radial particle current $q^1 \equiv q^r$ 
         [cf. Eq.~(\ref{heat-app})]
         \beq\label{current-radial}
              q^1 \equiv q^r=-\kappa\,\left(\frac{\pad f}{\pad r}+
                \frac{\gamma^2\,\Omega^2\,r}{c^2}\frac{(p'^0)^2}{p'}
                \frac{\pad f}{\pad p'}\right)\,.
         \eeq

        In the steady-state the fifth term in brackets of Eq.~(\ref{webb-app}) 
        becomes
         \beq
            \nablaa \, q^{\alpha} =\frac{1}{r}\frac{\pad}{\pad r}(r\,q^r)-
            \kappa\,(1+\gamma^2 v_z^2/c^2)\frac{\pad^2 f}{\pad z^2}\,,
         \eeq noting that the third component of the heat flux is given by
         \beq
            q^3=-\kappa\,(1+\gamma^2 v_z^2/c^2) \frac{\pad f}{\pad z}\,.
         \eeq 
         Finally, for the fourth term in brackets of Eq.~(\ref{webb-app}) we 
         have
         \beq
            u^{\alpha} \nablaa f =(\gamma v_z/c) \frac{\pad f}{\pad z}\,.
         \eeq 
   
        Now, by collecting together all the relevant terms of 
        Eq.~(\ref{webb-app})and by introducing a source term $Q$ (which may 
        depend on $r$,$z$ and $p'$), we finally arrive at the relevant 
        relativistic steady-state transport equation in cylindrical coordinates
        being appropriate for the considered rotating and shearing jet 
        flows:\\ \\
        \fbox{\parbox{8.5cm}{
        \beqn\label{webb-cylinder-app}
            &&\hspace{-0.3cm}
                \frac{1}{p'^2}\,\frac{\pad}{\pad p'}
                \left[p'\,(p'^{\,0})^2\, \frac{\gamma^2\,\Omega^2\, r}{c^2} 
                q^r 
                \right.\nonumber \\  
            &&\hspace{1.0cm} 
                \left. -\frac{\gamma^4\,r^2}{15}(1-v_z^2/c^2)
                \left(\frac{d\Omega}{dr}\right)^2\,p'^4 
                \,\tau_c\, \frac{\pad f}{\pad p'} \right]  
                \nonumber  \\
            &&\hspace{-0.2cm} +\,\,\gamma\,v_z\,\frac{\pad f}{\pad z}
                -\kappa (1+\gamma^2 v_z^2/c^2) \frac{\pad^2 f}{\pad z^2}
                +\frac{1}{r}\frac{\pad}{\pad r}(r\,q^r)=Q\,.
               \nonumber \\
        \eeqn}}\vspace{0.2cm}

      For purely azimuthal, special relativistic flows with $v_z=0$, the
      transport equation~(\ref{webb-cylinder-app}) reduces to Eq.~(5.2) 
      derived in WJM~94.\\

      As suggested by WJM~94, the derived transport equation may be cast  
      in a more suitable form by introducing the variable 
      \beq
         \Phi=\ln(H)
      \eeq replacing the comoving particle momentum variable $p'$.       
      Following WJM~94, we may define $H$ such that $(\pad f/\pad r)_H=
      -q^r/\kappa$, with $q^r$ given by Eq.~(\ref{current-radial}), the index 
      $H$ denoting a derivative at constant $H$, i.e.
      \beq\label{ham-app}
       H=p'^0\,c\,\exp\left(-\int^r \md r'\,\frac{\gamma^2\,\Omega^2\,r'}{c^2}
       \right)\,.
      \eeq
      With respect to a physical interpretation of $H$ we would like to note, 
      that in the case of rigid rotation (i.e. $\Omega=const.$) $H$ could be 
      related to the Hamiltonian for a bead on a rigidly rotating wire 
      [cf. WJM~94; see also Eq.~(\ref{hami-rigid})].\\      
      Hence, by writing $f(r,z,p')\rightarrow f(r,z,\Phi)$, the 
      relevant derivatives transform like
      \beqn\label{deri1}     
      \left(\frac{\pad f(r,z,\Phi)}{\pad r}\right)_{\Phi} 
           &=& \frac{\pad f(r,z,p')}{\pad r}
            + \left(\frac{\pad p'}{\pad r}\right)_{\Phi} 
              \frac{\pad f(r,z,p')}{\pad p'}
              \nonumber \\
           &=& -\frac{q^r}{\kappa}\,,
      \eeqn using $\pad p'/\pad p'^0=p'^0/p'$ and noting that
      $p'^0\,c=\exp[\int \md r'\,\gamma^2\,\Omega^2\,r'/c^2]\,\exp\Phi$. 
      As usual, the index $\Phi$ in Eq.~(\ref{deri1}) denotes a derivative 
      at constant $\Phi$. Similarly, for the momentum-derivatives we have
      \beqn\label{deri2}
      \frac{\pad f(r,z,p')}{\pad p'}=\frac{\pad \Phi}{\pad p'}
      \frac{\pad f(r,z,\Phi)}{\pad \Phi}=\frac{p'}{(p'^0)^2} 
      \frac{\pad f(r,z,\Phi)}{\Phi}
      \eeqn
      and consequently
      \beqn\label{deri3}
      \frac{\pad^2 f(r,z,p')}{\pad p'^2}
         &=&\frac{(p'^0)^2-2\,p'^2}{(p'^0)^4}
             \, \frac{\pad f(r,z,\Phi)}{\pad \Phi}
         \nonumber\\ 
         &&\,+\, \frac{p'^2}{(p'^0)^4}
            \,\frac{\pad^2 f(r,z,\Phi)}{\pad \Phi^2}\,.
      \eeqn

      Now, using Eq.~(\ref{deri1}) and collecting all terms together in 
      the transport equation~(\ref{webb-cylinder-app}) which depend on 
      $\pad f/\pad r$ and additionally recalling that $\kappa=
      (v'^2\,\tau_c)/3$ we may arrive at
      \beqn\label{collect1}
        &&\hspace{-0.65cm}
          \left[\frac{1}{r}+\frac{\pad \kappa/\pad r}{\kappa}\right]
          \! \left(\frac{\pad f}{\pad r}\right)_{\Phi}
          \!\!\!\! +\frac{1}{p'^2}
          \frac{\pad}{\pad p'}\left[p'\,(p'^0)^2\,
          \frac{\gamma^2\,\Omega^2\,r}{c^2}\right] 
          \!\! \left(\frac{\pad f}{\pad r}\right)_{\Phi}
          \nonumber\\
        &=& \frac{1}{r}\,(1+\beta)
            \left(\frac{\pad f}{\pad r}\right)_{\Phi} 
             \!\! +\,[3+ \alpha]\left(\frac{p'^0}{p'}\right)^2\,
             \frac{\gamma^2\,\Omega^2\,r}{c^2}
             \left(\frac{\pad f}{\pad r}\right)_{\Phi}
       \nonumber \\
      \eeqn where the position and momentum dependence of the collision
      time $\tau_c$ (and thus of the diffusion coefficient) has been
      caught into the definition of the variables $\alpha$ and $\beta$,
      i.e.  $\alpha$ and $\beta$ are given by
      \beq\label{alpha-beta-app}
       \alpha =\frac{\pad \ln \tau_c}{\pad \ln p'}\
       \qquad \mbox{\rm and} \qquad
       \beta = \frac{\pad \ln \tau_c}{\pad \ln r}\,,
      \eeq respectively.
      Eq.~(\ref{collect1}) has been obtained noting that 
      $v'^2 = c^2\,p'^2/(p'^0)^2$ and $[\pad \tau_c/\pad p']/\tau_c
      =\alpha/p'$ and using
      \beq
      \frac{\pad \kappa}{\pad p'}=\frac{1}{3}\left(\frac{\pad v'^2}{\pad p'}
      \,\tau_c + v'^2 \frac{\pad \tau_c}{\pad p'}\right)\,.
      \eeq
      In a similar manner, the terms depending on $\pad f/\pad p'$
      in Eq.~(\ref{webb-cylinder-app}) may be rewritten using 
      Eq.~(\ref{deri2}) as 
      \beqn
         &&\hspace{-0.55cm}
         \frac{1}{p'^2}\left[\frac{\pad}{\pad p'}\left(\frac{\gamma^4\,r^2}{15}
         \,(1-v_z^2/c^2)\,\left(\frac{\md \Omega}{\md r}\right)^2 \!
         p'^4\,\tau_c \right)\right]\frac{\pad f(r,z,p')}{\pad p'}
         \nonumber\\
         &=&\kappa\,\frac{\gamma^4\,r^2}{5\,c^2}\,(1-v_z^2/c^2)\,[4+\alpha]
         \left(\frac{\md \Omega}{\md r}\right)^2\,
         \frac{\pad f(r,z,\Phi)}{\pad \Phi}\,.
         \nonumber\\
      \eeqn

      Likewise, for the terms in Eq.~(\ref{webb-cylinder-app}) which depend
      on $\pad^2 f/\pad p'^2$ one finds 
      \beqn
        && \hspace{-0.55cm}
           \frac{1}{p'^2}\left(\frac{\gamma^4\,r^2}{15\,c^2}\,
            (1-\frac{v_z^2}{c^2})
           \left(\frac{\md \Omega}{\md r}\right)^2 p'^4\,\tau_c \right)
           \frac{\pad^2 f(r,z,p')}{\pad p'^2}
           \nonumber \\
       &=& \kappa\,\frac{\gamma^4\,r^2}{5\,c^2}\,(1-\frac{v_z^2}{c^2})
           \!\left(\frac{\md \Omega}{\md r}\right)^2 \left([1-2\,
           \frac{p'^2}{(p'^0)^2}] \,\frac{\pad f(r,z,\Phi)}{\pad \Phi}
           \right.
           \nonumber \\
       && \hspace{3cm} +\left.
           \frac{p'^2}{(p'^0)^2}\,\frac{\pad^2 f(r,z,\Phi)}{\pad \Phi^2}
           \right)\,
      \eeqn where Eq.~(\ref{deri3}) has been used.\\
      Now, collecting all relevant terms together the steady state transport
      equation~(\ref{webb-cylinder-app}) may finally be rewritten as \\ \\
      \fbox{\parbox{8.6cm}{
      \beqn\label{webb-transport-app}
          && \hspace{-0.45cm}
             \frac{\pad^2 f}{\pad r^2}+\left(\frac{1+\beta}{r}+[3+\alpha]\,
             \frac{\gamma^2\,\Omega^2\,r}{c^2}\,
             \left(\frac{p'^0}{p'}\right)^2 \right)\,\frac{\pad f}{\pad r}
             \nonumber \\
          &+&\frac{\gamma^4\,r^2}{5\,c^2}\,(1-v_z^2/c^2)
             \left(\frac{\md \Omega}{\md r}\right)^2
             \left(\left[5+\alpha-2\left(\frac{p'}{p'^0}\right)^2\right]
             \frac{\pad f}{\pad \Phi}\right.
             \nonumber\\
          && \hspace{4.6cm}\left.
             +\left(\frac{p'}{p'^0}\right)^2
             \frac{\pad^2 f}{\pad \Phi^2}\right)
             \nonumber\\
          &-& \frac{\gamma\,v_z}{\kappa}\,\frac{\pad f}{\pad z}
             + (1+\gamma^2\,v_z^2/c^2)\,\frac{\pad^2 f}{\pad z^2}
             =-\frac{Q}{\kappa}\,,
      \eeqn}} \\ \\
      with $f \equiv f(r,z,\Phi)$ and where the $r$-derivatives 
      should be understood as derivatives keeping $\Phi$ constant, i.e.
      we may treat $r,z,\Phi$ as independent variables. 
      Again, for $v_z=0$ the partial differential equation 
      (\ref{webb-transport-app}) reduces to the steady state transport equation
      given in WJM~94 [e.g. see their Eq.~(5.5)].\\
      For the present application where we are interested in highly 
      relativistic particles with $p'^0 \simeq p'$, the steady-state transport 
      equation~(\ref{webb-transport-app}) simplifies to\\ \\
      \fbox{\parbox{8.6cm}{
      \beqn\label{transport-simple-app}
      \frac{\pad^2 f}{\pad r^2}&+&\left(\frac{1+\beta}{r}+[3+\alpha]\,
            \frac{\gamma^2\,\Omega^2\,r}{c^2}\right)\,\frac{\pad f}{\pad r}
      \nonumber \\
        &+&\frac{\gamma^4\,r^2}{5\,c^2}\,(1-v_z^2/c^2)\!
           \left(\frac{\md \Omega}{\md r}\right)^2
           \left([3+\alpha]\frac{\pad f}{\pad \Phi}+
           \frac{\pad^2 f}{\pad \Phi^2}\right)
      \nonumber\\
        &-&\frac{\gamma\,v_z}{\kappa}\,\frac{\pad f}{\pad z}
         + (1+\gamma^2\,v_z^2/c^2)\,\frac{\pad^2 f}{\pad z^2}
         =-\frac{Q}{\kappa}\,.
      \eeqn}}

\section{Green's solutions for the steady state diffusive particle transport 
         equation}
      For general applications one may search for (two-dimensional) Green's 
      solutions of the steady state transport equation (see also WJM~94), i.e. 
      solutions of Eq.~(\ref{transport-simple-app}) with source term
      \beq\label{source}
         Q=q_0\,\delta(r-r_s)\,\delta(p'-p_s')\,\delta(z-z_s)\,,
      \eeq or equivalently (i.e. utilizing the properties of the delta 
      function) with source term
      \beq\label{source-term}     
         Q=\frac{q_0}{p_s'}\,\delta(r-r_s)\,\delta(\Phi-\Phi_s)\,\delta(z-z_s)\,,
      \eeq describing monoenergetic injection of particles with momentum
      $p'=p_s'$ at position $r=r_s$, $z=z_s$. 
      For consistency, the constant $q_0$ in Eqs.~(\ref{source}) and
      (\ref{source-term}) has to be defined such, that the relevant expression 
      satisfies the requirement that $\delta(\vec{r}-\vec{r}_s)\,
      \delta(\vec{p}\,'-\vec{p}_s\,')$ vanishes unless $r=r_s$, $\phi=\phi_s$, 
      $p'=p_s'$ and integrate to unity (or $N_s$ if $N_s$ particles are 
      injected) over all space and momentum directions. 
      Using cylindrical coordinates this requires
      \beq\label{source-q0}
         q_0=\frac{N_s}{8\,\pi^2\,p_s'^2\,r_s}\,.      
      \eeq
      Solutions of the transport equation~(\ref{transport-simple-app}) may 
      then be found by applying Fourier techniques, i.e. by using the double 
      Fourier transform defined by
      \beqn\label{fourier-transform}
         F(r;\mu,\nu)=\int_{-\infty}^{\infty}\!\!\! dz
                         \int_{-\infty}^{\infty}\!\!\! d\Phi\;
                         \exp[i\,(\nu\,\Phi+\mu\,z)]\;f(r,z,\Phi)\,,
         \nonumber\\
      \eeqn 
      where the inverse Fourier transform is given by
      \beqn\label{fourier-invers}
         f(r, z, \Phi)=\frac{1}{4\pi^2}
                 \int_{-\infty}^{\infty}\!d\mu\;\int_{-\infty}^{\infty}
                 \!d\nu\;\exp[-i\,(\nu\,\Phi+\mu\,z)]
                 \nonumber\\
                 \times F(r;\mu,\nu)\,.\,
      \eeqn 
      Denoting the Fourier transform of the source term $-Q/\kappa$ by 
      $\tilde{Q}$, we have
      \beqn
         \tilde{Q}&=&-\int_{-\infty}^{\infty}\!dz \int_{-\infty}^{\infty}\!
                    d\Phi\; \exp[i\,(\nu\,\Phi+\mu\,z)] \frac{q_0}{\kappa\,p_s'}
                    \nonumber \\
                  && \hspace{2cm}\times\,\delta(r-r_s)\,\delta(\Phi-\Phi_s)\,
                     \delta(z-z_s)
                    \nonumber\\
                  &=&-\frac{q_0}{\kappa_s\,p_s'}\,
                    \exp[i\,(\nu\,\Phi_s+\mu\,z_s)]\,\delta(r-r_s)\,.
      \eeqn
      Taking the Fourier transform of the transport 
      equation~(\ref{transport-simple-app}) one arrives at
      \beqn\label{fourier-transport}
         \frac{\pad^2 F}{\pad r^2}&+&\left(\frac{1+\beta}{r}+[3+\alpha]\,
          \frac{\gamma^2\,\Omega^2\,r}{c^2}\right)\,\frac{\pad F}{\pad r}
          \nonumber\\
          &-&\left[\frac{\gamma^4\,r^2\,(1-\frac{v_z^2}{c^2})}{5\,c^2}
           \left(\frac{\md \Omega}{\md r}\right)^2
           \left([3+\alpha]\,i\,\nu+\nu^2\right)\right] F
           \nonumber\\
           &-&\left[\frac{\gamma\,v_z}{\kappa}\,i\,\mu
           + (1+\gamma^2\frac{v_z^2}{c^2})\,\mu^2\right] F
           =\tilde{Q}\,.
      \eeqn
       Let $F_G$ be the (two-dimensional) Green's solution of this equation 
       satisfying homogeneous, i.e. zero Dirichlet boundary conditions, then 
       Fourier inversion [i.e. Eq.~(\ref{fourier-invers})] yields the required 
       Green's function solution $f_G(r,z,p';r_s,z_s,p_s')$.        
       Generally, the considered Fourier techniques yield the Green's function 
       for infinite domains. The proper Green's function for bounded domains 
       (i.e. for finite $z$) may be obtained, for example, by the method of 
       images (e.g. Morse \& Feshbach~\cite{morsefeshbach}, pp. 812-816) in 
       the case where the boundaries are restricted to straight lines in two 
       dimensions or planes in three dimensions. The steady state version
       of the Green's formula (cf. WJM~94, Eqs. [3.24],[7.12]) may then be 
       applied in order to arrive at the general solution of the considered
       transport equation.\\
       Yet, even for the simplification $v_z=0$ and $\gamma=const$
       (i.e. using a simple galactic rotation law $\Omega \propto 1/r$), the 
       Green's function solution for the steady state transport equation 
       Eq.~(\ref{transport-simple-app}) is not straightforward to evaluate 
       (see WJM~94).
       However, as we are primarily interested in an analysis of the azimuthal 
       effects of particle acceleration in rotating jet flows, we may be content 
       with a $z$-independent solution of the transport equation, i.e. with
       an investigation of the so-called one-dimensional Green's function which 
       preserves much of the physics involved (cf. Morse \& 
       Feshbach~\cite{morsefeshbach}, pp. 842-847). 
       This (one-dimensional) Green's function may be found by integrating the  
       two-dimensional ring source $Q$ (which depends on the space coordinates 
       $r_s, z_s$) from $z_s=-\infty$ to $z_s=\infty$; i.e. the one-dimensional
       Green's function represents the Green's solution for a steady state 
       monoenergetic injection of particles from a infinite cylindrical surface
       parallel to the jet axis at radius $r_s$. Hence, by utilizing the fourier 
       integral theorem it could be directly shown, that the one-dimensional Green's 
       function $f_{G,1D}$ corresponds to a $z$-independent solution of the transport 
       equation~(\ref{transport-simple-app}) with source term 
       $Q(r,p')=q_0\;\delta(r-r_s)\;\delta(p'-p_s')$, i.e. the (one-dimensional) 
       Green's function we are seeking for in the following analysis, represents 
       the solution of the modified transport equation\\ \\ 
       \fbox{\parbox{8.6cm}{
       \beqn\label{transport-one-dimensional}
         \frac{\pad^2 f}{\pad r^2}&+&\left(\frac{1+\beta}{r}+[3+\alpha]\,
            \frac{\gamma^2\,\Omega^2\,r}{c^2}\right)\,\frac{\pad f}{\pad r}
         \nonumber \\
           &+&\frac{\gamma^4\,r^2}{5\,c^2}\,(1-v_z^2/c^2)\!
           \left(\frac{\md \Omega}{\md r}\right)^2
           \left([3+\alpha]\frac{\pad f}{\pad \Phi}+
           \frac{\pad^2 f}{\pad \Phi^2}\right)
         \nonumber \\
         &=&-\frac{q_0}{\kappa\,p_s'}\,\delta(r-r_s)\,\delta(\Phi-\Phi_s)\,.
       \eeqn}}

\subsection{Rigid rotation profiles}
        In the case of solid body (uniform) rotation shearing in the 
        background flow is absent since the fluid moves without internal 
        distortions. 
        For $\Omega=\Omega_0=const.$, Eq.~(\ref{transport-one-dimensional}) 
        reduces to the purely spatial transport equation 
        \beqn\label{rigidtransport}
        \frac{\pad^2 f}{\pad r^2}+\left[\frac{1+\beta}{r}+ (3+\alpha)
        \frac{\tilde{\Omega}_0^2\,r/c^2}{(1-\tilde{\Omega}_0^2\,r^2/c^2)}
             \right]\,
             \frac{\pad f}{\pad r}=Q_0\,,
             \nonumber\\ 
         \eeqn where the constant $\tilde{\Omega}_0$ is defined by
         \beq\label{omega-tilde}
         \tilde{\Omega}_0=\frac{\Omega_0}{\sqrt{1-v_z^2/c^2}}\,, 
         \eeq while $Q_0= - q_0\,\delta(r-r_s)\,\delta(\Phi-\Phi_s)/
         (\kappa_s \,p_s')$, with $q_0$ given by Eq.~(\ref{source-q0}), and 
         where the diffusion coefficient is assumed to be of the form 
         \beq
             \kappa=\kappa_{\rm o}\,\left(\frac{p'}{p_{\rm o}'}\right)^{\alpha}
                     \left(\frac{r}{r_{\rm o}}\right)^{\beta}\,.
         \eeq with $\kappa_{\rm o}, p_{\rm o}'$ and $\alpha, \beta$ as constants,
         cf. Eqs.~(\ref{diffusion-coefficient}),~(\ref{alpha-beta-app}).\\
         For rigid rotation Eq.~(\ref{ham-app}) yields
         \beq\label{hami-rigid}
             H(r,p')= p'^0\,c\,\,(1-\Omega_0^2\,r^2/c^2-v_z^2/c^2)^{1/2}
                    = \frac{p'^0\,c}{\gamma}\,,
         \eeq where $\gamma$ denotes the Lorentz factor of the flow.
         As $H=H_s(r_s,p_s')$ is a constant of motion (cf. Noether's theorem, 
         see also WJM~94), the particle momentum $p'$ in the comoving frame 
         could be simply expressed as a function of the radial coordinate 
         \beq\label{energy-increase}
              p'(r)=m_0\,c\,\sqrt{\frac{H_s^2}{m_0^2\,c^4\,(1-
                            \Omega_0^2\,r^2/c^2-v_z^2/c^2)}-1}\,,
         \eeq with $m_0$ the rest mass of the particle.\\
         Basically, the solution space of the homogeneous part of 
         Eq.~(\ref{rigidtransport}) could be described by a set of two 
         independent solutions, e.g. by the functions $y_1(r)$ and $y_2(r)$ 
         with Wronskian $W(r)\equiv W(y_1(r),y_2(r))=y_1\,\md y_2/\md r
         - y_2\,\md y_1/\md r$, where for an appropriate choice $y_2(r)\equiv 
         1$. The relevant analytical expressions might be directly written down 
         for some special cases of interest: a.) For a constant diffusion 
         coefficient, i.e. $\alpha=\beta=0$, two independent solutions are given
         by $y_2(r) \equiv 1$ and 
         \beqn
          y_1(r)&=&\sqrt{1-\tilde{\Omega}_0^2\,r^2/c^2}\,
                  \left(\frac{4}{3}-\frac{\tilde{\Omega}_0^2\,r^2}{3\,c^2}\right)
                     \nonumber\\
                &&-\ln\frac{c\,(1+\sqrt{1-\tilde{\Omega}_0^2\,r^2/c^2})}
                     {\tilde{\Omega}_0\,r}\,
         \eeqn where for the Wronskian one simply finds
         \beq\label{wronski1}
                  W(r)=-r^{-1}\,
                  \left(1-\frac{\tilde{\Omega}_0^2\,r^2}{c^2}\right)^{3/2}\,.
         \eeq
         b.) In the case, where $\beta$ is negative, the solution $y_1(r)$ could 
         be expressed in terms of the incomplete Beta function (cf. Abramowitz \& 
         Stegun~\cite{abramstegun}, p.~263), i.e. one finally may arrive at the 
         system $y_2(r) \equiv 1$ and
         \beqn
          y_1(r)&=&\frac{1}{2}\left(\frac{\tilde{\Omega}_0}{c}\right)^\beta \times
                  B\left(\frac{\tilde{\Omega}_0^2\,r^2}{c^2}, -\frac{\beta}{2},
                  \frac{5+\alpha}{2}\right) \nonumber\\
                &&\hspace{4cm}{\rm for}\;\;\alpha>-5,\,{\rm and}\;\beta<0\, 
                   \nonumber\\
         \eeqn
         Note, that now the solutions $y_1, y_2$ have been defined such that the 
         appropriate Wronskian reduces to Eq.~(\ref{wronski1}) for 
         $\alpha=\beta=0$, i.e. we have  
         \beq
         W(r)=-r^{-(1+\beta)}\,
               \left(1-\frac{\tilde{\Omega}_0^2\,r^2}{c^2}\right)^{(3+\alpha)/2}\,.
         \eeq
         The general (one-dimensional Green's) solution of the inhomogeneous 
         differential equation Eq.~(\ref{rigidtransport}) with monoenergetic 
         source term $Q_0$ defined above could then be written as (e.g. Morse 
         \&  Feshbach~\cite{morsefeshbach}, p.~530)
         \beqn\label{general-solution}
            f(r,p')&=& y_1(r)\,
                   \left[k_1 -\int^r \frac{Q_0\,y_2(r)}{W(r)}\md r \right]
                   \nonumber \\
                    & &\hspace{1.2cm}
                    +y_2(r)\,\left[k_2 +\int^r \frac{Q_0\,y_1(r)}{W(r)}
                     \md r \right]\,,
         \eeqn where $k_1,\,k_2$ are integration constants specified by the 
         boundary conditions.\\
         In the disk-jet scenario the accretion disk is usually assumed to 
         supply the mass for injection into the jet, thus for simplicity one 
         may consider a rather hollow jet structure (cf. Marcowith et 
         al.~\cite{marcowithetal95}; Fendt~\cite{fendt97a}; Subramanian et 
         al.~\cite{subraetal99}) where the plasma motion in the azimuthal 
         direction is restricted to a region $r_{\rm in} \leq r \leq r_{\rm out} 
         < r_{\rm L}$ where $r_{\rm in}$ denotes the jet inner radius, 
         $r_{\rm out}$ the relevant outer radius and $r_{\rm L}$ the light 
         cylinder radius. 
         Particles are supposed to be injected at position $r_{\rm s}$ 
         with initial momentum $p_{\rm s}'$, where $r_{\rm in} < r_{\rm s} <
         r_{\rm out}$. By chosing homogeneous boundary conditions 
         $f(r = r_{\rm in})=0$ and $f(r = r_{\rm out})=0$, the integration 
         constants in Eq.~(\ref{general-solution}) are determined by
         \beqn
             k_1&=&-k_2\,\frac{1}{y_1(r_{\rm in})} \quad\; {\rm and}\nonumber\\
             k_2&=&\frac{\tilde{q}\,\,[y_1(r_{\rm out})-y_1(r_{\rm s})]}
                  {1-y_1(r_{\rm out})/y_1(r_{\rm in})}\,,
          \eeqn where 
          \beq\label{vorfaktor}       
          \tilde{q}= - \frac{q_0}{\kappa_s\,p_s'\,W(r_s)}\,\delta(\Phi-\Phi_s)\,,
          \eeq with $q_0$ given by Eq.~(\ref{source-q0}), i.e. 
          $q_0=N_s/(8\,\pi\,p_s'^2\,r_s)$.\\
          Therefore the (one-dimensional) Green's solution may be written as
          \beqn\label{insgesamt}
           f(r,p';r_s,p_s')= y_1(r)\,\left[k_1\,\theta(r-r_{\rm s})+k_1\,
                     \theta(r_{\rm s}-r) \right.
                 \nonumber\\
                 \hspace{-1cm}
                 \left. -\tilde{q}\,\theta(r-r_{\rm s})\right]
                 \nonumber\\                    
                 \quad
                 +\left[k_2\,\theta(r-r_{\rm s})+k_2\,\theta(r_{\rm s}
                 -r)+ \tilde{q}\,y_1(r_{\rm s})\,\theta(r-r_{\rm s})\right]\,,
                 \nonumber\\
          \eeqn where $\theta(x)$ denotes the Heaviside step function.
          The delta function in Eq.~(\ref{vorfaktor}) and Eq.~(\ref{insgesamt}) 
          indicates that the particle momentum in the comoving frame is directly 
          related to the relevant radial position by Eq.~(\ref{energy-increase}).
          In order to gain insight into the efficiency of the acceleration process
          one may introduce a spatial weighting function $N(r)$ defined 
          by 
          \beqn\label{rigid-final-solution}
               f(r,p';r_s,p_s')= N(r)\,\delta(\Phi-\Phi_s)
                                  = N(r)\,H_s\,\delta(H-H_s)\,. \nonumber\\
          \eeqn 

\subsection{Keplerian rotation profiles}
         In the case of Keplerian rotating background flow with $\Omega(r)=
         k\,r^{-3/2}$, $k=\sqrt{G\,M}$ generally both, shear and centrifugal
         acceleration, will occur. 
         By applying a simple Fourier transformation 
         (cf. Eqs.~[\ref{fourier-transform}],[\ref{fourier-invers}]), i.e. 
         \beq\label{fourier-transform-simple}
              F=\int_{-\infty}^{\infty}\md \Phi\, \exp[i\,\nu\,\Phi]\,f\,,
         \eeq where the inverse Fourier transform is given by
         \beq
              f=\frac{1}{2\,\pi} \int_{-\infty}^{\infty} \md \nu\, 
                   \exp[-i\,\nu\,\Phi]\,F\,,
         \eeq the transport equation~(\ref{transport-one-dimensional}) could be 
         written as
         \beqn\label{transport-keplerian-fourier}
            \frac{\pad^2 F}{\pad r^2}&+&
            \frac{1}{r}\,\left(a_1 + a_2 \frac{\tilde{\gamma}(r)^2}{r}\right)
            \frac{\pad F}{\pad r}\nonumber\\
            &-&\frac{\tilde{\gamma}(r)^4}{r^3}\left(i\,\nu\,a_3+a_4\,\nu^2\right) F
           =\tilde{Q}_0\,.
         \eeqn where $\tilde{\gamma}(r)$ is defined by $\tilde{\gamma}(r)=
         \gamma(r)\,\sqrt{1-v_z^2/c^2}$, with $\gamma$ the Lorentz factor of the 
         flow, and where $\tilde{Q}_0$ denotes the Fourier transform of the source
         term $Q_0=-q_0 \,\delta(r-r_s)\,\delta(\Phi-\Phi_s)\,/(\kappa_s \,p_s')$, i.e.
         \beq\label{source-fourier}
            \tilde{Q}_0=-\frac{q_0\,\exp[i\,\nu\,\Phi_s]}{\kappa_s\,p_s'}
                      \,\delta(r-r_s)\,.
         \eeq
         The abbreviations $a_1, a_2, a_3, a_4$ in 
         Eq.~(\ref{transport-keplerian-fourier}) are defined by
         \beqn\label{a_1}
            a_1&=&(1+\beta)\,,\\
            a_2&=&(3+\alpha)\,\frac{G\,M}{(c^2-v_z^2)}\,,\\
            a_3&=&\frac{9}{20}\,(3+\alpha)\,\frac{G\,M}{(c^2-v_z^2)}\,,\\
            a_4&=&\frac{9}{20}\,\frac{G\,M}{(c^2-v_z^2)}\,.\label{a_4}
         \eeqn
         An analytical evaluation of Eq.~(\ref{transport-keplerian-fourier}) 
         is rather complicated. However, a simple set of solutions may be 
         found in the case of $r$ being large such that the rotational velocity 
         becomes non-relativistic and the approximation $\tilde{\gamma}(r)=
         (1-G\,M/[(c^2-v_z^2)\,r])^{-1/2} \simeq 1 $ holds. For, the Fourier 
         transformed transport equation then simplifies to
         \beqn\label{transport-keplerian-simple}
            \frac{\pad^2 F}{\pad r^2}+
            \frac{1}{r}\,\left(a_1 + \frac{a_2}{r}\right)\frac{\pad F}{\pad r}
            -\frac{1}{r^3}\left(i\,\nu\,a_3+a_4\,\nu^2\right) F
            =\tilde{Q}_0\,,
            \nonumber\\        
         \eeqn which, using the substitution $y=a_2/r$, $a_2 \neq 0$ (i.e. $\alpha 
         \neq -3$), leads to
         \beq\label{Kummer}
             y\,\frac{\pad^2 F}{\pad y^2} + (2-a_1-y)\,\frac{\pad F}{\pad y}
             -\frac{i\,\nu\,a_3 + a_4\,\nu^2}{a_2}\,F=0,
         \eeq for the homogeneous part of Eq.~(\ref{transport-keplerian-simple}).
         Eq.~(\ref{Kummer}) is known in the literature as Kummer's equation
         (e.g. Abramowitz \& Stegun~\cite{abramstegun}, p.~504).
         For the general case where $a_2 \neq 0$ and $(2-a_1) \neq -n$, 
         $n \in N_0$, the complete solution of this equation, i.e. of
         the homogeneous part of Eq.~(\ref{transport-keplerian-simple}),
         may be written as
         \beq
            F_H(r,\nu) = c_1\,f_1(r,\nu) + c_2\,f_2(r,\nu)
         \eeq where the functions $f_1, f_2$ are given by
         \beqn\label{kepler-homogen}
            f_1(r,\nu) &=& M\left(\frac{i\,\nu\,a_3 + a_4\,\nu^2}{a_2},
                            2-a_1, \frac{a_2}{r}\right)\\            
            f_2(r,\nu) &=& U\left(\frac{i\,\nu\,a_3 + a_4\,\nu^2}{a_2},
                            2-a_1, \frac{a_2}{r}\right)\,.
         \eeqn 
         Here, $M(a,b,y)$ and $U(a,b,y)$ denote the confluent hypergeometric 
         functions (cf. Abramowitz \& Stegun~\cite{abramstegun}, pp.~504f; 
         Buchholz~\cite{buchholz}, pp.~1-9), with $M(a,b,y)$ being characterized 
         by the series representation
         \beqn
            M(a,b,y)&=&1+\frac{a}{1!\,b}\,y + \frac{a\,(a+1)}{2!\,b\,(b+1)}\,y^2
                     \nonumber\\  
                    & &+ \frac{a\,(a+1)\,(a+2)}{3!\,b\,(b+1)\,(b+2)}\,y^3                 
                     + \ldots
         \eeqn while $U(a,b,y)$ is given by the series 
         \beqn 
             U(a,b,y)&=&\frac{\pi}{\sin \pi b} \left(\frac{M(a,b,y)}
                     {\Gamma(1+a-b)\,\Gamma(b)} \right.
                     \nonumber\\
                     & &\hspace{1cm}
                     \left.- y^{1-b}\,\frac{M(1+a-b,2-b,y)}{\Gamma(a)\,\Gamma(2-b)}   
                     \right)\!,
         \eeqn with $\Gamma(x)$ the Gamma function.\\           
         $M(a,b,z)$ has a simple pole at $b=-n$ for $a \neq -m$ or for
         $a=-m$ if $m>n$, and is undefined for $b=-n$, $a=-m$ and $m \leq n$. 
         $U(a,b,z)$, however, is defined even for $b \rightarrow \pm n$.\\ 
         For the relevant Wronskian one has (e.g. Abramowitz \& Stegun
         \cite{abramstegun}, p.~505)
         \beq
            W(y)\equiv W\left(M(a,b,y),U(a,b,y)\right)=-\frac{\Gamma(b)
                 \,y^{-b}\,e^y}{\Gamma(a)}\,.
         \eeq 
         Thus, for the Wronskian $W(r)\equiv W(f_1,f_2)$ we find
         \beq
             W(r,\nu)=\frac{a_2}{r^2}\,
                  \Gamma(2-a_1)\,\left(\frac{a_2}{r}\right)^{-(2-a_1)}
                  \frac{e^{a_2/r}}{\Gamma(a(\nu))}\,,
         \eeq where 
         \beq\label{anu}
               a(\nu)=\frac{i\,\nu\,a_3 + a_4\,\nu^2}{a_2}\,.
         \eeq
         For the general solution of the inhomogeneous fourier 
         equation~(\ref{transport-keplerian-fourier}) one finds
         [cf. Eq.~(\ref{general-solution})]
         \beqn\label{fourier-solution}
            F(r,\nu) &=& f_1(r,\nu)\,\left[k_1\,\theta(r-r_{\rm s})+
                            k_1\, \theta(r_{\rm s}-r) \right.
                            \nonumber\\
                        & &\hspace{2.5cm}\left. -\tilde{q}(\nu)\, f_2(r_{\rm s},\nu)\,
                           \theta(r-r_{\rm s})\right]\nonumber\\
                        &+& f_2(r,\nu)\,\left[k_2\,\theta(r-r_{\rm s})+
                            k_2\, \theta(r_{\rm s}-r)\right.
                            \nonumber\\
                        & &\hspace{2.5cm}\left. +\tilde{q}(\nu)\,f_1(r_{\rm s},\nu)\,
                           \theta(r-r_{\rm s})\right]\,,\nonumber\\
         \eeqn where $\tilde{q}(\nu)$ has been defined by
         \beq
            \tilde{q}(\nu)= - q_0\,
                          \frac{\exp[i\,\nu\,\Phi_s]}{\kappa_s \,p_s'\,W(r_s,\nu)}\,,
         \eeq with $q_0$ given by Eq.~(\ref{source-q0}) and where $k_1,\,k_2$ are 
         integration constants specified by the boundary conditions. For homogeneous 
         Dirichlet conditions at the boundaries $r_{\rm in}$ and $r_{\rm out}$, i.e. 
         $F(r_{\rm in},\nu)=F(r_{\rm out},\nu)=0$ with $r_{\rm in} < r < r_{\rm out}$ 
         and $r_{\rm in} < r_s < r_{\rm out}$ the integration constants are fixed
         (for each $\nu$) and given by
         \beqn
           k_2(\nu) &=& -\tilde{q}(\nu)\,f_1(r_{\rm in},\nu)\,
                    \nonumber\\
                  & &\quad \times \frac{f_2(r_{\rm out},\nu)\,f_1(r_{\rm s},\nu)
                  -f_1(r_{\rm out},\nu)\,f_2(r_{\rm s},\nu)}
                  {f_2(r_{\rm out},\nu)\,f_1(r_{\rm in},\nu)-
                   f_1(r_{\rm out},\nu)\,f_2(r_{\rm in},\nu)}\,,\nonumber \\ \\
             k_1(\nu) &=&  \frac{f_2(r_{\rm in},\nu)}{f_1(r_{\rm in},\nu)}\;k_2\,.
          \eeqn
          By Fourier inversion, we now may obtain the required (one-dimensional) 
          Green's function
          \beq\label{integral-complex}
            f(r,\Phi; r_s,\Phi_s)=\frac{1}{2\,\pi}\,\int_{-\infty}^{\infty}\md \nu\,
                         \exp[-i\,\nu\,\Phi]\,F(r,\nu)\,,
          \eeq with $F(r,\nu)$ given by Eq.~(\ref{fourier-solution}).
          As may be obvious from the foregoing investigation this fourier 
          inversion is not easy to evaluate, not even using numerical methods. 
          However, in order to cope with the integration, we may consider the 
          following substitution ($\alpha \neq -3$)
          \beq
               \omega=\frac{3}{\sqrt{20\,(3+\alpha)}}\,
                      \left(\nu+\frac{3+\alpha}{2}\,i\right)\,, 
          \eeq for which one finds [cf. Eq.~(\ref{anu}) and 
          Eqs.~(\ref{a_1})-(\ref{a_4})]
          \beq
             a(\nu)=\omega^2 + \sigma^2\,,
          \eeq where $\sigma$ is defined by $\sigma=3\,\sqrt{3+\alpha}/\sqrt{80}$.\\ 
          Performing the substitution in Eq.~(\ref{integral-complex}),
          one arrives at an integral with the path of integration $A$  
          now in the complex plane, i.e. parallel to the real axis at a distance 
          $3\,(3+\alpha)\,i/[2\,\sqrt{20\,(3+\alpha)}\,]$, extending from $-R$ to 
          $R$ with $R \rightarrow \infty$. We may close the path by choosing a 
          rectangular contour $C$ which consists of the stated parallel $A$, the 
          real axis, and the outer lines $B_1$, $B_2$ parallel to the imaginary 
          axis at $R$ and $-R$, respectively. For the relevant ranges of $\alpha$ 
          ($\alpha \neq -3$; $2-a_1 \neq -n$) of interest, the integrand has no 
          poles within the region bounded by $C$ and thus, by virtue of Cauchy's 
          integral theorem, the value of the contour integral around $C$ sums to 
          zero. For $R \rightarrow \infty$ the integrals over $B_1$ and over $B_2$ 
          vanish as might be verified by using asymptotic expansion formulas for the 
          integrand. Noting that by means of Euler's equation $e^{i\,x}=\cos x + 
          i\,\sin x$ we have
          \beq
              \int_{-\infty}^{\infty} e^{-i\,x}\,f(x^2)\,\md x=
               2 \int_0^{\infty} f(x^2)\,\cos x \,\md x\,,
          \eeq and collecting all relevant expressions together, one 
          finally may arrive at the integral ($2-a_1 \neq -n; \alpha \neq -3$)
          \beqn\label{final-integral}
               f(r,p';r_{\rm s},p_s')&=&g(r_s,p_s',\alpha,\beta)\;
                   \exp\left[-\frac{3+\alpha}{2}(\Phi-\Phi_s)\right]\,\nonumber\\
                && \hspace{-2.5cm}\times \int_0^{\infty}\!\! \md \omega 
                   \cos\left(\frac{\sqrt{20\,(3+\alpha)}}{3}\,\omega\,(\Phi-\Phi_s)\right)
                   \times \Gamma(\omega^2+\sigma^2)\,
                   \nonumber\\
                && \hspace{-1cm}\times    
                   \left[f_1(r, \omega^2)\,f_2(r_{\rm out},\omega^2)\,
                             \frac{h(r_{\rm in}, r_{\rm s}; \omega^2)}
                              {h_N(r_{\rm in}, r_{\rm out};\omega^2)}\right.
                   \nonumber\\
                && \hspace{-0.7cm}
                   + \left. f_2(r, \omega^2)\,f_1(r_{\rm out},\omega^2)
                              \frac{h(r_{\rm s}, r_{\rm in};\omega^2)}
                              {h_N(r_{\rm in}, r_{\rm out};\omega^2)}\right]
                   \nonumber\\
                && \hspace{3cm} {\rm for}\quad r > r_{\rm s}  \\ \nonumber \\
                &=& g(r_s,p_s',\alpha,\beta)\;
                    \exp\left[-\frac{3+\alpha}{2}(\Phi-\Phi_s)\right]\,\nonumber\\
                &&  \hspace{-2.5cm}\times\int_0^{\infty}\!\! \md \omega 
                    \cos\left(\frac{\sqrt{20\,(3+\alpha)}}{3}\,\omega\,(\Phi-\Phi_s)\right)
                    \times \Gamma(\omega^2+\sigma^2)\,
                    \nonumber\\
                && \hspace{-1cm}\times
                   \left[f_1(r,\omega^2)\,f_2(r_{\rm in},\omega^2)\,
                             \frac{h(r_{\rm out}, r_{\rm s};\omega^2)}
                              {h_N(r_{\rm in}, r_{\rm out};\omega^2)}\right.
                   \nonumber\\
                && \hspace{-0.7cm}
                   + \left. f_2(r,\omega^2)\,f_1(r_{\rm in},\omega^2)
                              \frac{h(r_{\rm s}, r_{\rm out};\omega^2)}
                              {h_N(r_{\rm in}, r_{\rm out};\omega^2)}\right] 
                   \nonumber \\
                && \hspace{3cm} {\rm for}\quad r < r_{\rm s}                     
              \eeqn where we have introduced the following abbreviations
              \beqn
                 g(r_s,p_s',\alpha,\beta) &=& \frac{2\,q_0\,r_{\rm s}\sqrt{5 (3+\alpha)}}
                       {3\,\pi\,\kappa_s\,p_s'\,\Gamma(2-a_1)}
                       \left(\frac{a_2}{r_{\rm s}}\right)^{1-a_1}
                       \!\! e^{-a_2/r_{\rm s}}\,,
                       \nonumber\\  \\
                 f_1(r,\omega^2)&=&M(\omega^2+\sigma^2, 2-a_1,a_2/r)\,,\\
                 f_2(r,\omega^2)&=&U(\omega^2+\sigma^2, 2-a_1,a_2/r)\,,\\
                 h(r_{\rm in}, r_{\rm s};\omega^2)&=&
                     f_1(r_{\rm in},\omega^2)\,f_2(r_{\rm s},\omega^2)
                     \nonumber\\
                     &&\quad\quad\quad 
                       - f_2(r_{\rm in},\omega^2)\,f_1(r_{\rm s},\omega^2)\,,\\ 
                 h_N(r_{\rm in}, r_{\rm out};\omega^2)&=&
                     f_2(r_{\rm out},\omega^2)\,f_1(r_{\rm in},\omega^2)
                     \nonumber\\
                     &&\quad\quad\quad 
                     - f_1(r_{\rm out},\omega^2)\,f_2(r_{\rm in},\omega^2),
              \eeqn with $q_0$ given by Eq.~(\ref{source-q0}).
              Eq.~(\ref{final-integral}) may be evaluated using numerical methods 
              (cf. Wolfram~\cite{wolfram96}). $f_1(r,\omega^2)$ and 
              $f_1(r,\omega^2)$ are bounded for $\omega = 0$ while the integrand 
              behaves well enough when $\omega \rightarrow \infty$ to allow for a 
              numerical evaluation.

\subsection{Flat rotation profiles}
           In the case of flat (galactic-type) rotation with $\Omega(r)=
           \Omega_0\,r_0/r$, the co-operation of centrifugal and shear 
           effects could be analysed more directly. Applying Fourier 
           transformation (cf. Eq.~[\ref{fourier-transform-simple}])
           the transport equation~(\ref{transport-one-dimensional})
           simplifies to
           \beq\label{transport-flat-fourier}
               \frac{\pad^2 F}{\pad r^2} +\frac{a}{r}\,\frac{\pad F}{\pad r}
               +\frac{b(\nu)}{r^2}\,F=\tilde{Q}_0\,.
           \eeq where $a$ and $b(\nu)$ are defined by
           \beqn
                a &=& (1+\beta)+(3+\alpha)\,\gamma^2\, \eta^2\,, \\
                b(\nu) &=& \left[(3+\alpha)\,i\,\nu + \nu^2 \right]
                           \,\frac{\gamma^4\,\eta^2}{5}\,(1-v_z^2/c^2)
           \eeqn with $\gamma=1/\sqrt{1-\eta^2-v_z^2/c^2}$ the Lorentz 
           factor of the flow and $\eta \equiv v_{\theta}/c=\Omega_0\, r_0/c$ 
           its rotational velocity, and where $\tilde{Q}_0$ is given
           by Eq.~(\ref{source-fourier}). Solutions for the homogeneous part 
           of Eq.~(\ref{transport-flat-fourier}) could then be directly 
           written down, i.e. one has
           \beqn\label{flat-fourier-solutions}
             F_1(r,\nu)&=& r^{\frac{1}{2}\left(1-a-\sqrt{(a-1)^2+4\,b(\nu)}\right)}\,,\\
             F_1(r,\nu)&=& r^{\frac{1}{2}\left(1-a+\sqrt{(a-1)^2+4\,b(\nu)}\right)}\,,\\
           \eeqn with Wronskian $W(r,\nu)$ fixed by
           \beq
              W(r,\nu)=\frac{\sqrt{(a-1)^2 + 4\,b(\nu)}}{r^a}\,.
           \eeq
           The general solution of the inhomogeneous fourier 
           equation~(\ref{transport-flat-fourier}) in the case of homogeneous 
           Dirichlet conditions at the boundaries $r_{\rm in}$ and $r_{\rm out}$,
           may then be determined following the steps given in the previous 
           subsection, e.g. see Eqs.~({\ref{fourier-solution}) - 
           (\ref{integral-complex}). In order to calculate the inverse Fourier
           transform, it proves useful to apply the following substitutions
           \beqn
               \omega & = & \nu +\frac{1}{2}\,(3+\alpha)\,i \\
               \sigma & = & \frac{3+\alpha}{2}\,
                          \sqrt{1+\frac{5}{\gamma^2\,\eta^2\,(1-v_z^2/c^2)} 
                          \left(\frac{\beta}{3+\alpha}+\gamma^2\,\eta^2\right)^2}
                          \,,
                          \nonumber\\
           \eeqn for which one has
           \beq
              \frac{1}{2}\sqrt{(a-1)^2 + 4\,b(\nu)}=\frac{\gamma^2\,\eta}{\sqrt{5}}
                                                    \sqrt{1-v_z^2/c^2}\,
                                                    \sqrt{\omega^2 +\sigma^2}\,.
           \eeq
           The complete solution may then be found by proceeding as presented in 
           the previous subsection.

\end{document}